\documentclass[fleqn,usenatbib]{mnras}
\usepackage{newtxtext,newtxmath}

\usepackage[T1]{fontenc}

\DeclareRobustCommand{\VAN}[3]{#2}
\let\VANthebibliography\thebibliography
\def\thebibliography{\DeclareRobustCommand{\VAN}[3]{##3}\VANthebibliography}

\usepackage{graphicx}	
\usepackage{float}
\usepackage{amsmath}	
\usepackage{amssymb}	
\usepackage{tabularx}
\usepackage{lineno} 

\hypersetup{linkcolor=red,citecolor=blue,filecolor=cyan,urlcolor=magenta}

\newcommand{\rj}{ R_{\rm 0}}

\newcommand{\rhoj}{\rho_{0}}

\title[Generation of spine-sheaths and shear acceleration]{Particle acceleration in shearing flows: the self-generation of turbulent spine-sheath structures in relativistic MHD jet simulations}
\author[Wang et al.]{
Jie-Shuang Wang,$^{1}$\thanks{jiesh.wang@gmail.com}
Brian Reville,$^{1}$
Yosuke Mizuno,$^{2,3}$
Frank M. Rieger,$^{1,4}$
Felix A. Aharonian$^{1,5,6}$\\
$^{1}$Max-Planck-Institut f\"ur Kernphysik, Saupfercheckweg 1, D-69117 Heidelberg, Germany\\
$^{2}$Tsung-Dao Lee Institute, Shanghai Jiao Tong University, 520 Shengrong Road, Shanghai, 201210, China\\
$^{3}$School of Physics \& Astronomy, Shanghai Jiao Tong University, 800 Dongchuan Road, Shanghai, 200240, China\\
$^{4}$Institute for Theoretical Physics, University of Heidelberg, Philosophenweg 12, D-69120 Heidelberg, Germany\\
$^{5}$Dublin Institute for Advanced Studies, 31 Fitzwilliam Place, Dublin 2, Ireland\\
$^{6}$High Energy Astrophysics Laboratory, RAU, 123 Hovsep Emin St Yerevan 0051, Armenia
}

\date{}
\pubyear{}

\begin{document}

\pagerange{\pageref{firstpage}--\pageref{lastpage}}

\label{firstpage}

\maketitle

\begin{abstract}
X-ray observations of several kiloparsec-scale extragalactic jets favour a synchrotron origin. The short cooling times of the emitting electrons requires distributed acceleration of electrons up to sub-PeV energies. In a previous paper, we found that this can be self-consistently explained by a shear acceleration model, where particles are accelerated to produce power-law spectra with a spectral index being determined mainly by the velocity profile and turbulence spectrum. 
In this paper, we perform 3D relativistic magneto-hydrodynamic simulations to investigate the formation of a spine-sheath structure and the development of turbulence for a relativistic jet propagating into a static cocoon. 
We explore different spine velocities and magnetic field profiles with values being chosen to match typical Fanaroff–Riley type I/II jets.
We find that in all cases a sheath is generated on the interface of the spine and the cocoon mainly due to the Kelvin-Helmholtz
instability. The large scale velocity profile in the sheath is close to linear. 
Turbulence develops in both the spine and the sheath, with a turbulent velocity spectrum consistent with Kolmogorov-scaling. 
The implications for shear particle acceleration are explored, with a focus on the particle spectral index. 
\end{abstract}


\section{Introduction}\label{section:intro}

Jets of active galactic nuclei (AGN) are observed on all scales extending from the vicinity of the black-hole event-horizon up to megaparsec distances \citep[see][for a review]{Blandford2019}. 
In large-scale AGN jets, bright diffuse X-ray emission has been observed out to distances of several
hundreds of kiloparsec \citep[e.g.][]{Harris2006}. 
Numerous observations in recent years support a synchrotron origin of X-rays in nearby large-scale jets, such as optical polarimetry measurements \citep[e.g.][]{Cara2013,Georganopoulos2016,Perlman2020} and GeV gamma-ray observation of some high-power FR II jets 
\citep[e.g.][]{Meyer2014,Meyer2015,Breiding2017}. A synchrotron X-ray origin is certain in the case of the low-power Fanaroff–Riley (FR) I 
radio galaxy Centaurus A, as the same electrons produce the diffuse TeV emission \citep{HESS2020}.

A synchrotron origin of X-rays requires sub-PeV electrons in these large-scale jets, i.e. typical synchrotron photon energy is
$E_{\rm syn}=2 ({E_e/{\rm 0.1~PeV}})^2({B/10~\mu{\rm G}})$ keV, where $B$ is the magnetic 
field strength.
On the other hand, the synchrotron cooling time of a relativistic electron with energy $E_{\rm e}$ is $t_{\rm cool}=1.25({E_{\rm e}/{\rm 0.1PeV}})^{-1}
({B/10\mu{\rm G}})^{-2}~{\rm kyr}$, corresponding to a maximum cooling length of $\sim0.4$\,kpc
decreasing further if the magnetic fields are higher.
The observed X-ray structures of the large-scale jets, have sizes exceeding 
kpc \cite[e.g.][]{Harris2006}, indicating that an \textit{in-situ} (re-)acceleration mechanism is 
required. Shear particle acceleration can serve as a natural explanation \citep{Liu2017,Wang2021MNRAS,Tavecchio2020}.
In our previous work \citep{Wang2021MNRAS}, we obtained an exact solution of the steady-state 
Fokker–Planck equation for shear acceleration in the presence of synchrotron losses, and found 
that the particle distribution resembles a power-law with an exponential-like cutoff.
The corresponding synchrotron radiation from such a population of electrons was shown to reproduce 
the observed optical-to-X-ray spectral energy distributions (SEDs) of Centaurus A and 3C 273. 

Shear acceleration is naturally expected in fast, velocity-shearing flows.
It can be understood as a second-order-Fermi-type acceleration mechanism, where particles 
can gain energy by scattering on the turbulence embedded in the velocity-shear flows 
\citep{Berezhko1981,BerezhkoKrymskii1981,Earl1988,Rieger2004,Rieger2007,Rieger2016,Webb2018,Webb2019,Rieger2019Galax}.
The resulting accelerated particle spectrum depends mainly on the velocity profile and turbulence spectrum 
\citep{Liu2017,Webb2018,Webb2019,Webb2020,Rieger2019ApJ,Rieger2021,Rieger2022ApJ,Wang2021MNRAS}.
Observationally, the existence of such velocity-shear layers (or sheaths) has been also 
supported by high-resolution radio imaging and polarization studies of large-scale jets 
\citep[e.g.,][]{Laing2014,Gabuzda2014,Nagai2014,Boccardi2016,Blandford2019}.
However, the detailed properties of such velocity-shearing sheaths remain to be studied.

One way of doing this, is through relativistic magneto-hydrodynamic (RMHD) simulations.
RMHD simulations have been extensively conducted to study the dynamics of relativistic jets \citep[see][for a recent review]{Marti2019Galax}.
The interaction between a jet and its ambient medium has been explored to study the global jet stability and morphology, where the formation of a hot cocoon and the development of MHD instabilities can clearly be observed \citep[e.g.][]{Marti1997ApJ,Komissarov1999MNRAS,Scheck2002MNRAS,Perucho2007MNRAS,Rossi2008A&A,Mignone2010MNRAS,Tchekhovskoy2016MNRAS,Rossi2017A&A,Perucho2019MNRAS,Mukherjee2020MNRAS}. 
Typically, in these simulations the numerical resolution in the jet region is highly limited due to the large simulation domain, and this is usually inadequate to follow accurately the MHD instabilities.

Dedicated analytical and numerical studies have also been performed to study the properties of MHD instabilities in jets. Two instabilities of importance are the Kelvin-Helmholtz instability (KHI), which can take place with or without magnetic fields near the interface of the velocity shear region  \citep[e.g.][]{Ferrari1978A&A,Hardee1979ApJ,Birkinshaw1991MNRAS,Baty2002ApJ,Perucho2004A&A,Mizuno2007ApJ,Bodo2013MNRAS,Bodo2019MNRAS,Sironi2021ApJL,Borse2021A&A,Chow2022arXiv},
and the current-driven instability (CDI), which takes place in the presence of toroidal magnetic fields \citep[e.g.][]{Lyubarskii1999MNRAS,Nakamura2007ApJ,Hardee2007ApJ,Mizuno2009ApJ,Mizuno2012ApJ,Mizuno2014ApJ,Kim2018,Bromberg2019ApJ,Ortuno-Macias2022ApJ}, though see also \citet{2018NatAs...2..167G}.
In general, CDI dominates in high-magnetization jets, while KHI dominates in low-magnetization jets, 
though the interaction between the CDI and KHI may also provide a stabilizing effect in low-magnetization jets \citep[e.g.][]{Baty2002ApJ}. 
In addition, kinetic plasma simulations are utilised to investigate both the jet dynamics and the particle acceleration through the dissipation of magnetic energy \citep[e.g.][]{Sironi2021ApJL,Ortuno-Macias2022ApJ}.
It has been suggested that these accelerated particles can serve as seed particles for shear acceleration \citep[e.g.][]{Wang2021MNRAS}.

In this paper, we investigate the self-generation of a sheath by simulating a relativistic spine jet propagating into a static cocoon using three-dimensional RMHD simulations. We study the resultant jet properties including the developed turbulence and possible shear acceleration. 
In Section \ref{sec:simulation}, we introduce the parameters and initial conditions for our simulations, and show the results on the jet dynamics. 
In Section \ref{sec:shear}, we analyse the velocity and turbulence properties, and discuss the implications for shear acceleration.
Conclusions and discussion are presented in Section \ref{sec:conclusion}.

\section{Numerical Setup and the Jet Dynamics}\label{sec:simulation}

\subsection{Numerical Setup}
We perform a numerical study of relativistic magnetised jet-spine flows propagating within a surrounding 
cocoon in order to investigate the properties of the self-generated spine-sheath structure. 
The PLUTO code \citep{Mignone2007ApJS} is employed to solve the RMHD equations in three-dimensional Cartesian geometry.
We adopt a Taub-Mathews equation of state \citep{Mathews1971ApJ,Mignone2005ApJS,Mignone2007ApJS}, where 
the specific enthalpy is $h=5\Theta/2+\sqrt{9\Theta^2/4+1}$ and $\Theta=p_g/\rho c^2$ represents the dimensionless temperature, with $p_{g}$ and $\rho$ being the thermal gas' pressure and density respectively.

The jet is initialised as a cylindrical flow with a spine radius ($\rj$) and its symmetry axis parallel to the $y$ axis. 
The boundary conditions are periodic in the $y$-direction mimicking a jet of infinite extent, with outflow boundary conditions used at the other simulation boundaries. 
The jet rest-mass density is $\rhoj=n_0m_p$ with $n_0= 10^{-6}$ and $m_p$ being the proton mass. 
The jet propagates into a hot stationary cocoon, which is formed with a density $\rho_{\rm c}=2\rhoj$ \citep[e.g.][]{Mukherjee2020MNRAS} due to the interaction between the jet and the ambient medium.
The density profile in the cocoon ($r>R_0$) is smoothed according to $\rho=(\rho_{\rm c}-\rhoj)\tanh[(r/\rj-1)/0.01] +\rhoj$. 
The velocity is only assigned within $\rj$ to ensure energy is only injected in this region.

To facilitate our study of the turbulent jet dynamics, we introduce small transverse velocity field perturbations on the jet-cocoon interface following the approach of \cite{Rossi2008A&A},
\begin{equation}
\left(\beta_x, \beta_z\right) = \frac{\tilde{A}}{24} \sum_{m=0}^{2} \sum_{l=1}^{8} \cos(m\phi + k_l y)(\cos\phi,\sin\phi),\label{eq:v_xz}
\end{equation}
with 
\[k_l = \frac{(0.5, 1, 2, 3, 0.03, 0.06, 0.12, 0.25)}{\beta_0\sqrt{1/\Upsilon \Theta +1/(\Upsilon -1)}}\] and $\Upsilon=13/9$. 
The transversal component of velocity is $\beta_{x,z}=v_{x,z}/c$.
The azimuth is defined as $\phi = \tan^{-1}(z/x)$. 
The perturbation amplitude is $\tilde{A} =\sqrt{\epsilon+0.5\epsilon^2}/\Gamma_0(1 + \epsilon)$, where we set $\epsilon = 0.001$ for all simulations.

A helical magnetic field configuration is initialised in both the spine and the cocoon. 
The field consists of an axial component $B_y$ and a toroidal component $B_\phi$, where
\begin{eqnarray}
B_y=\frac{B_1}{1+(r/a)^2},\nonumber\\
B_\phi=\frac{B_2(r/a)}{1+(r/a)^2},\label{eq:b_y,t}
\end{eqnarray}
with $a=\rj/2$ \citep[e.g.,][]{Mizuno2009ApJ}. This choice results in a constant pitch parameter $\Pi=rB_y/B_\phi=aB_1/B_2$. 
The normalisation parameters $B_1$ and $B_2$ determine the mean magnetization parameter of the axial and toroidal components,
\begin{equation}
    \sigma_{y,\phi}=< B_{y,\phi}^2>/8\pi \rhoj c^2,\label{eq:sigma}
\end{equation}
with the mean value given by $<f>=\int_0^{\rj} 2\pi f r dr/\int_0^{\rj} 2\pi r dr$. 
The magnetization inside the spine is then $\sigma=\sigma_y+\sigma_\phi$.

The magnetic field profile matches a screw pinch configuration, for which the equilibrium condition is
\begin{equation}
     B_\phi\frac{\partial (rB_\phi)}{\partial r}+ r B_y\frac{\partial B_y}{\partial r}+4 \pi r\frac{\partial p_{\rm g}}{\partial r}=0.\label{eq:p_balance}
\end{equation}
The gas pressure can be shown to satisfy 
\begin{equation}
     p_{\rm g}=p_{\rm g,0}+\frac{B_1^2-B_2^2}{8\pi} \left( 1-\frac{1}{(1+(r/a)^2)^2}\right),\label{eq:p_g}   
\end{equation}
where $p_{\rm g,0}=\Theta_0 \rhoj c^2$ is the gas pressure at $r=0$.

The initial values of jet velocity $(\beta_y=v_y/c=\beta_0)$, magnetization ($\sigma$), and temperature ($\Theta_0$) for each cases are listed in Table \ref{tab:1}. 
We choose to study three initial velocities with $\beta_0=0.6$ (labeled V6) for typical FR~I jets and $\beta_0=0.9, ~0.99$ (labeled V9 and V99) for FR~II jets. 
The corresponding kinetic luminosity ($L_{\rm K}=\pi \rj^2  v_y\Gamma_0^2  \rhoj h c^2$) is also shown in Table \ref{tab:1}, where $\Gamma_0$ is the jet bulk Lorentz factor. 
To explore the effect of magnetic field, we study three magnetization values with $\sigma_y=10^{-1},~10^{-2},~10^{-3}$ (labeled B-1, B-2, and B-3) and an axial/toroidal field dominated case (V6BA-2/V6BT-2 case). 
According to Eq. \ref{eq:sigma}, the jet average magnetic field is $\bar{B}_0=\sqrt{2\sigma B_0}=194\sqrt{\sigma}\mu$G in our simulations with $B_0=\sqrt{4\pi \rhoj c^2}=137\mu$G.

The jets are assumed to be cold with $\Theta_0=0.01-0.09$, where we adopt higher temperatures for magnetically dominated jets to avoid negative pressure (see Eq. \ref{eq:p_g}).
The averaged magnetic pressure in the spine is $p_B=\sigma \rhoj c^2$, which along with Eq. (\ref{eq:sigma}) indicates $p_B/p_g=\sigma/\Theta$. 
Therefore the jet pressure is comparable with the gas pressure when $\sigma\sim\Theta_0$. 
Thus, in our setup of $\Theta_0= (0.01-0.09)$, $\sigma=0.02$ implies the magnetic-pressure is comparable to the gas pressure, while $\sigma=0.2$ and $\sigma=0.002$ means magnetic or gas-pressure dominated respectively. 

It has been found that for cold jets, KHI dominates in low-magnetisation jets and CDI can dominate in high-magnetisation jets \citep{Bodo2013MNRAS}. 
In our simulations, we also find that CDI is dominant in the V6BT-2 and V9B-1 cases with a low-temperature $\theta_0\lesssim0.05$, leading to a strong deceleration of the jet spine. 
However, as we are mainly interested in the development of the sheath caused by the KHI, we adopt a slightly higher temperature ($\theta_0=0.09$) for the V6BT-2 and V9B-1 cases to avoid the domination of CDI.

\begin{table*} 
    \begin{tabular}{ c|c|c|c|c|c|c|c |c }
    \hline
Runs$^*$ &$\beta_0$  & $\sigma_y $ &$\sigma_\phi$ & Box size & Grid points & $\Theta_0 $ & $\rj$ & $L_{\rm K}$(erg/s)\\\hline

V6B-1 & 0.6  & $10^{-1}$ &$10^{-1}$ & $6.0\rj$ & $375^3$ &  $0.01$ &0.1kpc & $1.3\times10^{43}$ \\
V6B-1-SB & 0.6  & $10^{-1}$ &$10^{-1}$ & $4.8\rj$& $300^3$ &  $0.01$ &0.1kpc & $1.3\times10^{43}$ \\
V6B-1-LR & 0.6  & $10^{-1}$ &$10^{-1}$ & $6.0\rj$& $200^3$ &  $0.01$ &0.1kpc & $1.3\times10^{43}$ \\ \hline 

V6B-2 & 0.6  & $10^{-2}$ &$10^{-2} $ & $6.0\rj$& $375^3$ &  $0.01$ &0.1kpc & $1.3\times10^{43}$ \\
V6BA-2 & 0.6  & $0.016$ &$0.004$ & $6.0\rj$& $375^3$ &  $0.01$ &0.1kpc & $1.3\times10^{43}$ \\
V6BT-2 & 0.6  & $0.004$ &$0.016$ & $6.0\rj$& $375^3$ &  $0.09$ &0.1kpc & $1.6\times10^{43}$ \\ \hline 

V6B-3 & 0.6  & $10^{-3}$ &$10^{-3}$ & $6.0\rj$& $375^3$ &  $0.01$  &0.1kpc & $1.3\times10^{43}$\\\hline \hline 

V9B-1 & 0.9  & $10^{-1}$ &$10^{-1} $ & $8.0\rj$ & $500^3$ &  $0.09$ & 1\,kpc & $6.7\times10^{45}$ \\
V9B-2 & 0.9  & $10^{-2}$ &$10^{-2} $ & $8.0\rj$ & $500^3$  &  $0.04$ & 1\,kpc & $7.0\times10^{45}$\\
V9B-3 & 0.9  & $10^{-3}$ &$10^{-3} $ & $8.0\rj$ & $500^3$  &  $0.02$ & 1\,kpc & $6.7\times10^{45}$\\ \hline 
V99B-2 & 0.99  & $10^{-2}$ &$10^{-2} $ & $8.0\rj$ & $500^3$  &  $0.07$ & 1\,kpc & $7.9\times10^{46}$\\
\hline
    \end{tabular}
    \caption{The initial parameters used in the simulations. 
    The grid size is fixed at $0.016\rj$ except for V6B-1-LR case, which uses a lower resolution (LR). We also test the effect of box size by setting a smaller box (SB) length in the V6B-1-SB case. \\
    $^*$Terminology: V stands for the jet spine velocity, BA/BT/B indicates that the magnetic field is dominated by the axial/toroidal/both component.}
    \label{tab:1}
\end{table*}

\subsection{Results on Jet Dynamics}

\begin{figure}
    \centering
    \includegraphics[width=0.48\textwidth]{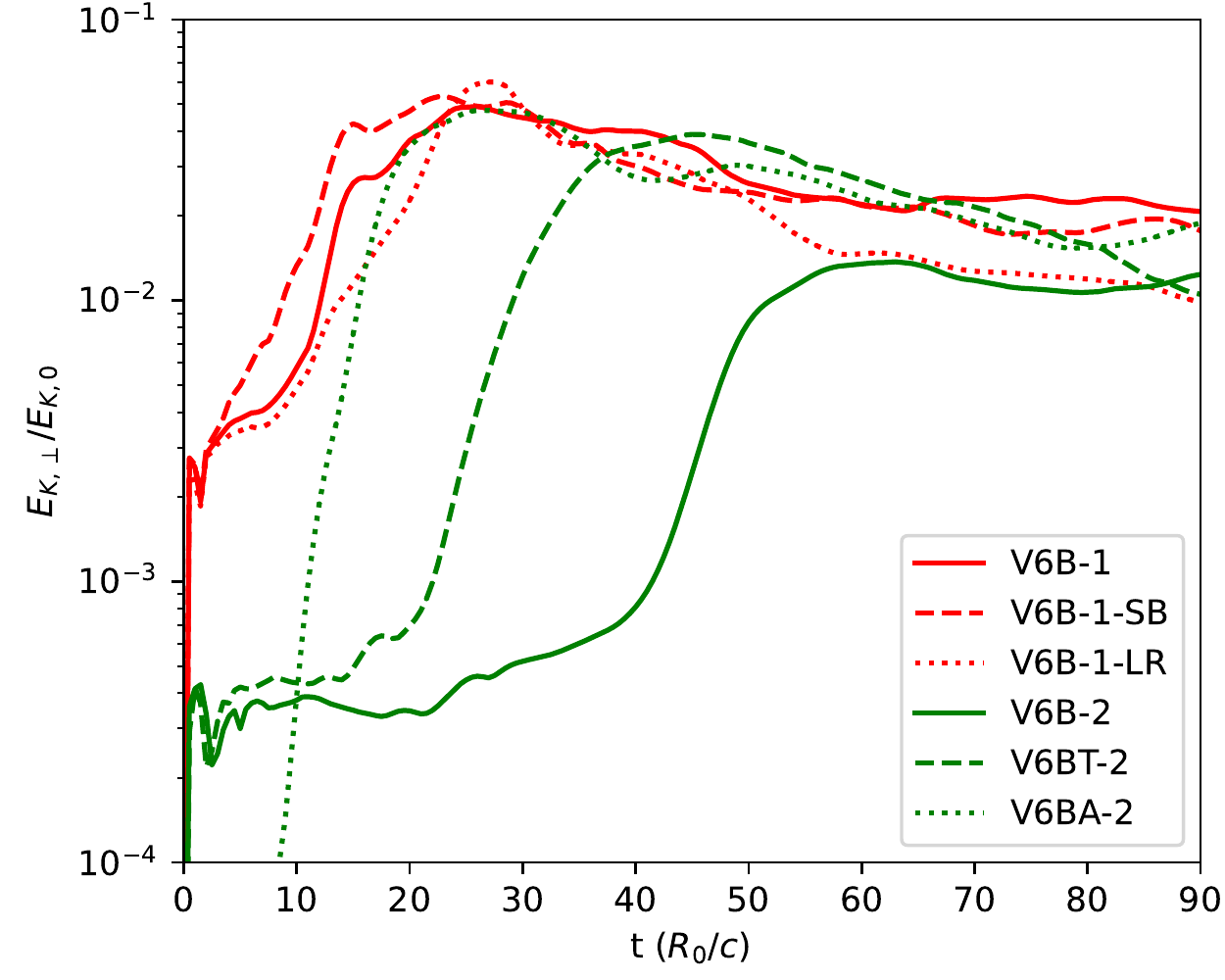}
    \caption{ The non-axial component of the kinetic energy $E_{K, \perp}\equiv \rho c^2(\beta_x^2+\beta_z^2)/2$ is used to indicate the development of KHI. The kinetic energy is defined in the Newtonian approximation and is normalised to the initial kinetic energy $E_{K,0}$. We test the effects of simulation box size and resolution with the cases V6B-1, V6B-1-SB, and V6B-1-LR. The V6B-2, V6BT-2, and V6BA-2 cases are used to study the effect of the magnetic field configurations. }
    \label{fig:test}
\end{figure}

\begin{figure}
    \centering
    \includegraphics[width=0.48\textwidth]{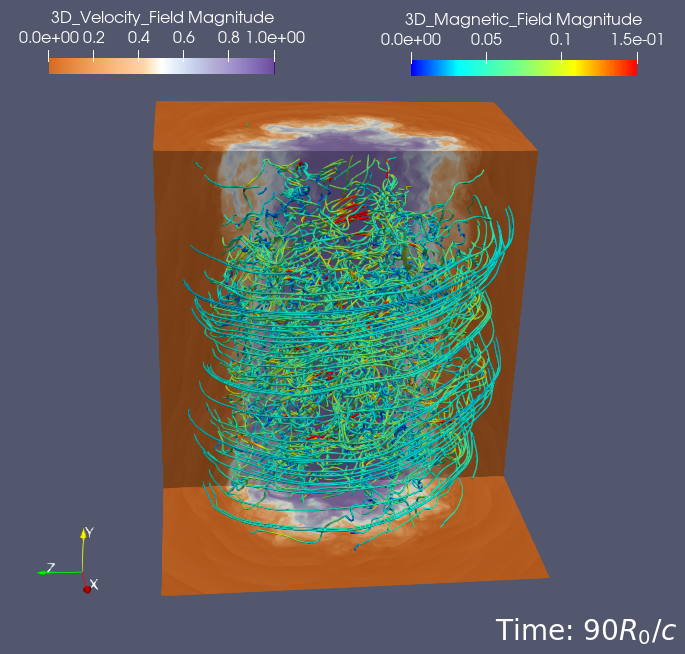}
    \caption{ An example of the simulation results, the 3D velocity ($\beta$) magnitude distribution and the magnetic field-lines ($\vec{B}$) in 3D, are shown for the V9B-3 case at $t=90\rj/c$. The magnetic field is normalised to $B_0=\sqrt{4\pi \rhoj c^2}=137\mu$G. }
    \label{fig:V9B-3map}
\end{figure}

The global behaviour of the time evolution of the simulations are similar within different cases. 
Due to the application of strong toroidal fields in Cartesian geometry, the initial configuration is not in exact pressure balance, resulting in a transient short-lived period of relaxation.
Following this, the small perturbations imposed on the transition layer between the jet and the cocoon lead to the growth of KHI driven by the velocity shear in this layer. 
Fig.~\ref{fig:test} shows the time evolution of the non-axial component of the kinetic energy, $E_{K,\perp} \equiv \rho c^2(\beta_x^2+\beta_z^2)/2$, which are used to indicate the development of KHI.
In linear stage, we see the exponential growth of non-axial component of the kinetic energy, and large eddies are gradually developed at the transition layer. 
Turbulence developed by eddies will amplify the magnetic field locally. 
In the saturation stage, the system tends to evolve non-linearly. 
The CDI may also grow in this stage and help to saturate KHI \citep{Baty2002ApJ}.
The pressure imbalance at the transition layer will affect the jet and the cocoon in this stage, which can generate shocks and waves. 
The transition layer becomes fully turbulent at the later stage of the simulations.
An example of the velocity and magnetic field structure at this stage is shown in Fig. \ref{fig:V9B-3map}. These are further explored in Section \ref{sec:turbulence}.

To assess the validity of the results on the large scale dynamics that follow, including the influence of the boundaries as well as grid resolution,
we explored different box sizes ($4.8\rj, 6.0\rj$) and grid resolutions ($0.016\rj, 0.03\rj$).
The case V6B-1 is the reference case where we use a box size of $6.0\rj$ and a grid resolution of $0.016\rj$ (see Table \ref{tab:1}). 
In the cases V6B-1-SB or V6B-1-LR, a smaller box size or a lower grid resolution is used, respectively.
We find that the box size has a negligible effect on the development of KHI provided it exceeds the width of the generated spine-sheath structure (see Section \ref{sec:shear} for more details). 
The resolution is seen to have an effect on both the growth and saturation stage of the KHI. 
In the lower resolution case, the growth rate is slightly smaller and $E_{K,\perp}$ is mildly suppressed in the saturation stage. 
Thus in the following work, we adopt a high resolution with a grid size of $0.016\rj$ and a box size of $6 \rj$ for FR I jets ($\beta_0=0.6$) and $8 \rj$ for FR II jets ($\beta_0\geq0.9$), corresponding to uniformly spaced computational domains with $375^3$ cells and $500^3$ cells respectively.

\begin{figure}
    \centering
    \includegraphics[width=0.48\textwidth]{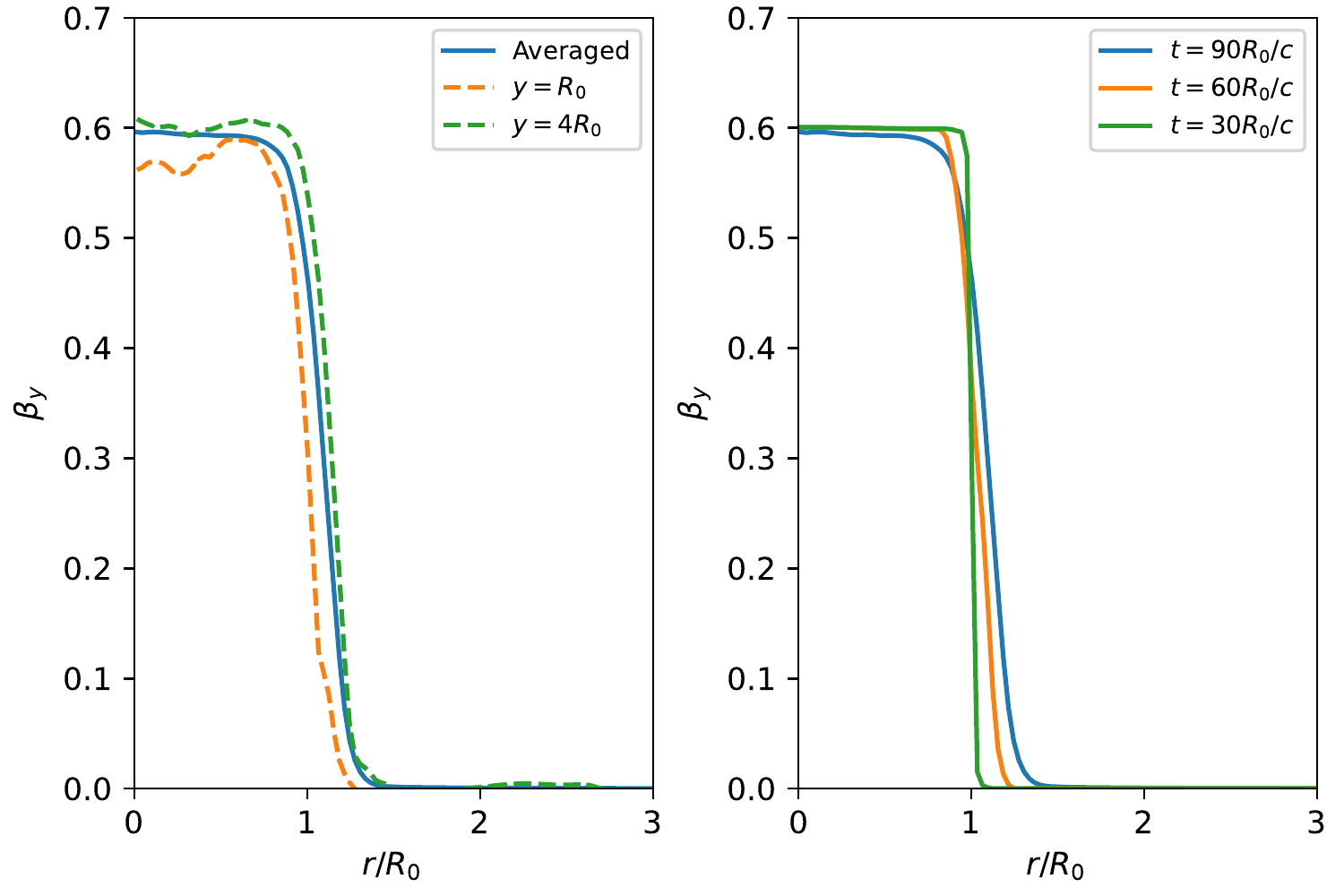}
    \caption{The azimuthally averaged axial velocity profiles for the V6B-2 case as a function of radius. 
    Left panel: Profiles at fixed time $t=90\rj/c$. The solid line shows the axial velocity averaged in the $y$ direction over the full length of the box. For comparison, we show examples of the profiles at two fixed points along the jet, at $y=1 \& 4\,\rj$.  
    Right panel: We show the $y$-averaged velocity profiles at different simulation time. 
    }\label{fig:V6B-2-y-time}
\end{figure}

We examine the axial velocity profile as a function of radius to measure the spine-sheath structure. 
In the left panel of Fig. \ref{fig:V6B-2-y-time}, we show the velocity profiles of the V6B-2 case at different locations along $y$. 
It is evident that a shearing layer is formed between the jet and the cocoon due to the KHI; a spine-sheath structure is generated.
Since the velocity profile varies along the jet axis we use the averaged velocity profile to represent the generated spine-sheath structure. 
The sheath is defined as the layer in which the velocity decreases continuously, which is approximately in the range $\sim0.8\rj-1.3\rj$ as shown in Fig. \ref{fig:V6B-2-y-time}.
The spine is the region enclosed by the sheath. 
In the right panel we show the averaged profile at different simulation times. 
The velocity profile at the early stage of the KHI development ($t=30\rj/c$) differs only slightly from the initial setup, and the jet width continues to expand in the saturation stage ($t=60,~90\rj/c$), indicating the location and extent of the sheath.

\begin{figure}
    \centering
    \includegraphics[width=0.48\textwidth]{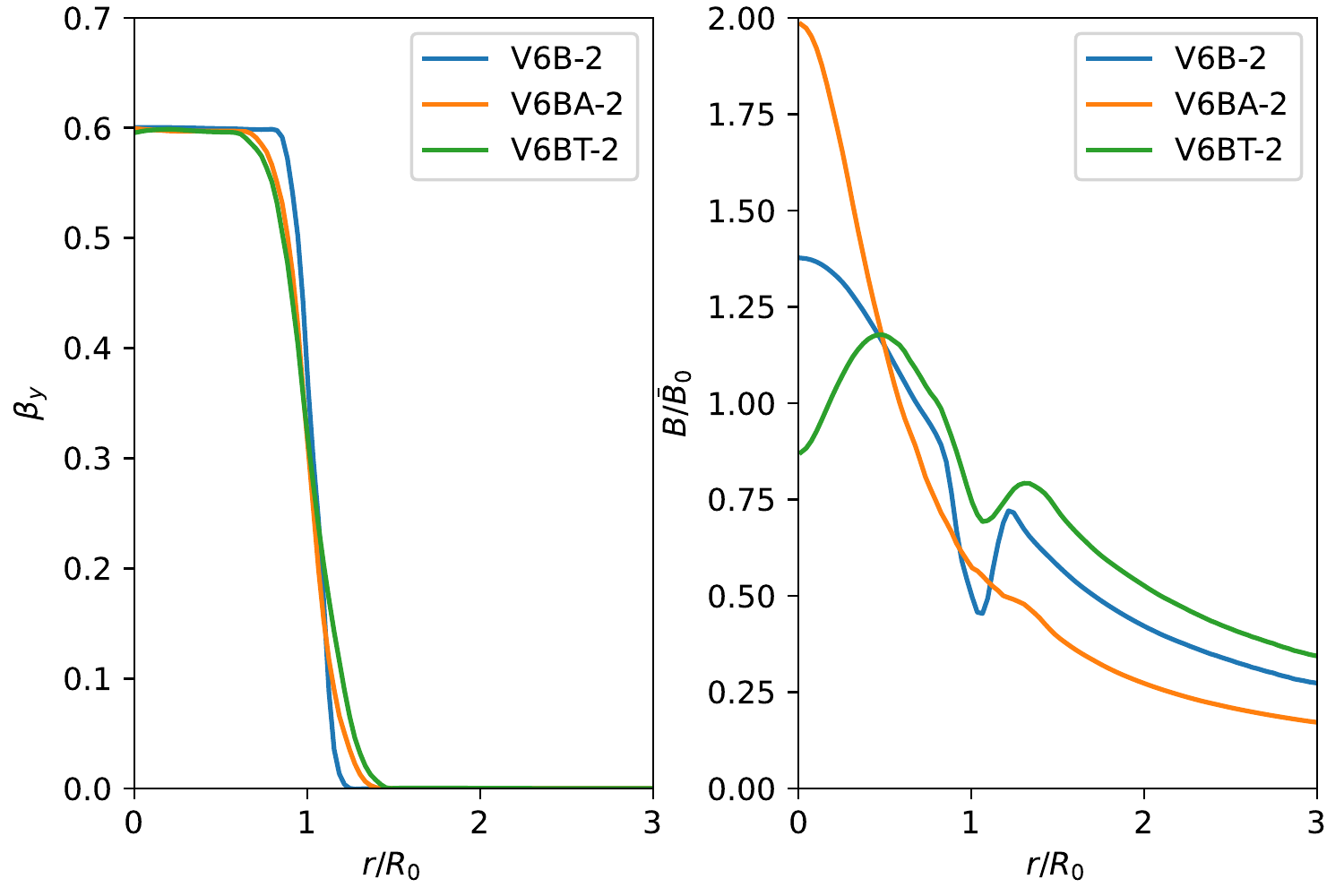}
    \caption{The azimuthally and $y$-averaged velocity (left panel) and magnetic field (right panel) profiles for the cases V6B-2 at $t=60\rj/c$, V6BT-2 at $t=40\rj/c$ and V6BA-2 at $t=20\rj/c$.}
    \label{fig:BTP}
\end{figure}

We simulate three different magnetic configurations for FR I type jets with $\sigma=0.02$ to investigate the effect of magnetic field configuration: an equal-field case (V6B-2 with $\sigma_y=\sigma_\phi$), an axial-field-dominated case (V6BA-2) and a toroidal-field-dominated case (V6BT-2), where the temperature $\Theta_0$ is the same for V6B-2 and V6BA-2 cases, and a higher temperature is assigned for V6BT-2 to avoid negative pressure (see Eq.~\ref{eq:p_g}). 
Fig.~\ref{fig:test} shows that the KHI grows with a different growth rate in different cases. 
This is because a high toroidal field \citep{Baty2002ApJ} or a low temperature \citep{Hardee2007ApJ} tends to suppress KHI.  
The V6B-2 and V6BT-2 cases experience longer relaxation times due to the larger applied toroidal fields.
In the early saturation stage shown in Fig.~\ref{fig:BTP}, the velocity and magnetic field profiles deviate slightly from the initial setup.
Fig.~\ref{fig:BTP} also shows that the V6BA-2 and V6BT-2 cases have slightly wider sheaths than the V6B-2 case. 
However, the difference is insignificant in the velocity profiles. We thus fix $\sigma_y=\sigma_\phi$ in the following studies.

\begin{figure}
    \centering
    \includegraphics[width=0.48\textwidth]{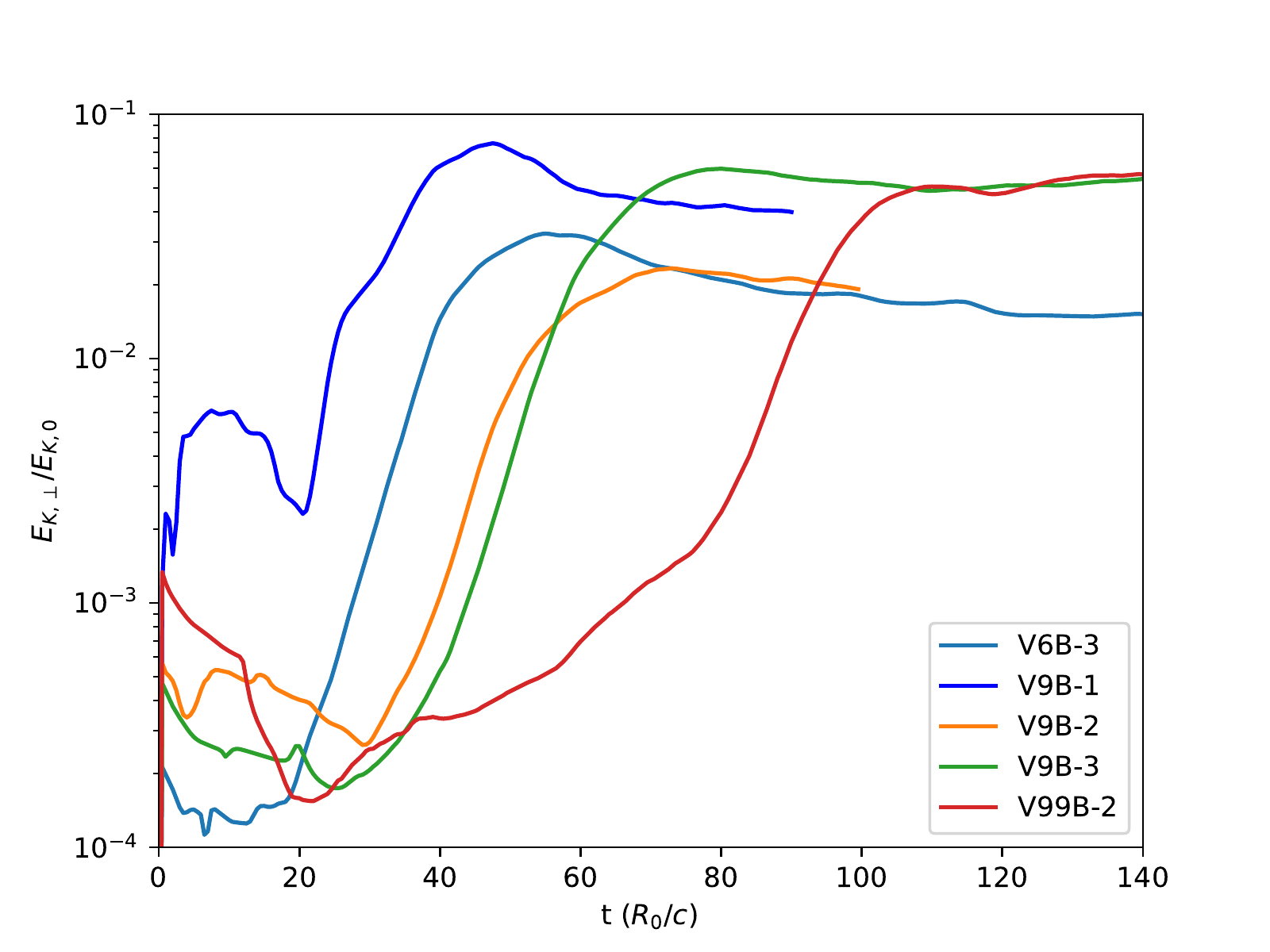}
    \caption{Plot showing, similar to  Fig. \ref{fig:test}, the non-axial component of the kinetic energy $E_{K, \perp}\equiv \rho c^2 (\beta_x^2+\beta_z^2)/2$ to the initial kinetic energy $E_{K,0}$ to indicate the growth rate of KHI for the cases V6B-3, V9B-1, V9B-2, V9B-3, and V99B-2.}
    \label{fig:v69dy}
\end{figure}

We consider three magnetisation scenarios with $\sigma_y=\sigma_\phi=10^{-1},~10^{-2},~10^{-3}$ for both FR I ($\beta_0=0.6$) and FR II ($\beta_0=0.9$) type jets to explore the impact of magnetisation. 
We also study one more highly relativistic case - V99B-2 with $\beta_0=0.99$ and $\sigma_y=\sigma_\phi=10^{-2}$ to investigate the effect of spine velocity. 
The evolution of KHI in these jets is shown in Fig. \ref{fig:test} and \ref{fig:v69dy}. 
We simulated these jets up to the late saturation stage to determine the spine-sheath properties and their implication for shear acceleration. 

\section{Results and Implications on Shear Acceleration}\label{sec:shear}

Shear acceleration can be understood as a stochastic type Fermi acceleration mechanism, where particles are accelerated via scattering off the turbulence embedded in the shearing flows. 
In a previous paper \citep{Wang2021MNRAS}, we studied the steady-state Fokker–Planck equation for shear acceleration and obtained an exact solution for the particle spectrum. 
The spectrum consists of a power-law component with an exponential-like cutoff, where the power-law index is determined by the turbulence spectral index and the velocity profile \citep[Eqs. 10, 11 and 3 in][]{Wang2021MNRAS}. 
In this paper, we study such properties for simulated jets with self-generated spine-sheath structures.

\subsection{Velocity and Magnetic-field Profiles}\label{sec:profile-vB}

\begin{figure}
    \centering
    \includegraphics[width=0.48\textwidth]{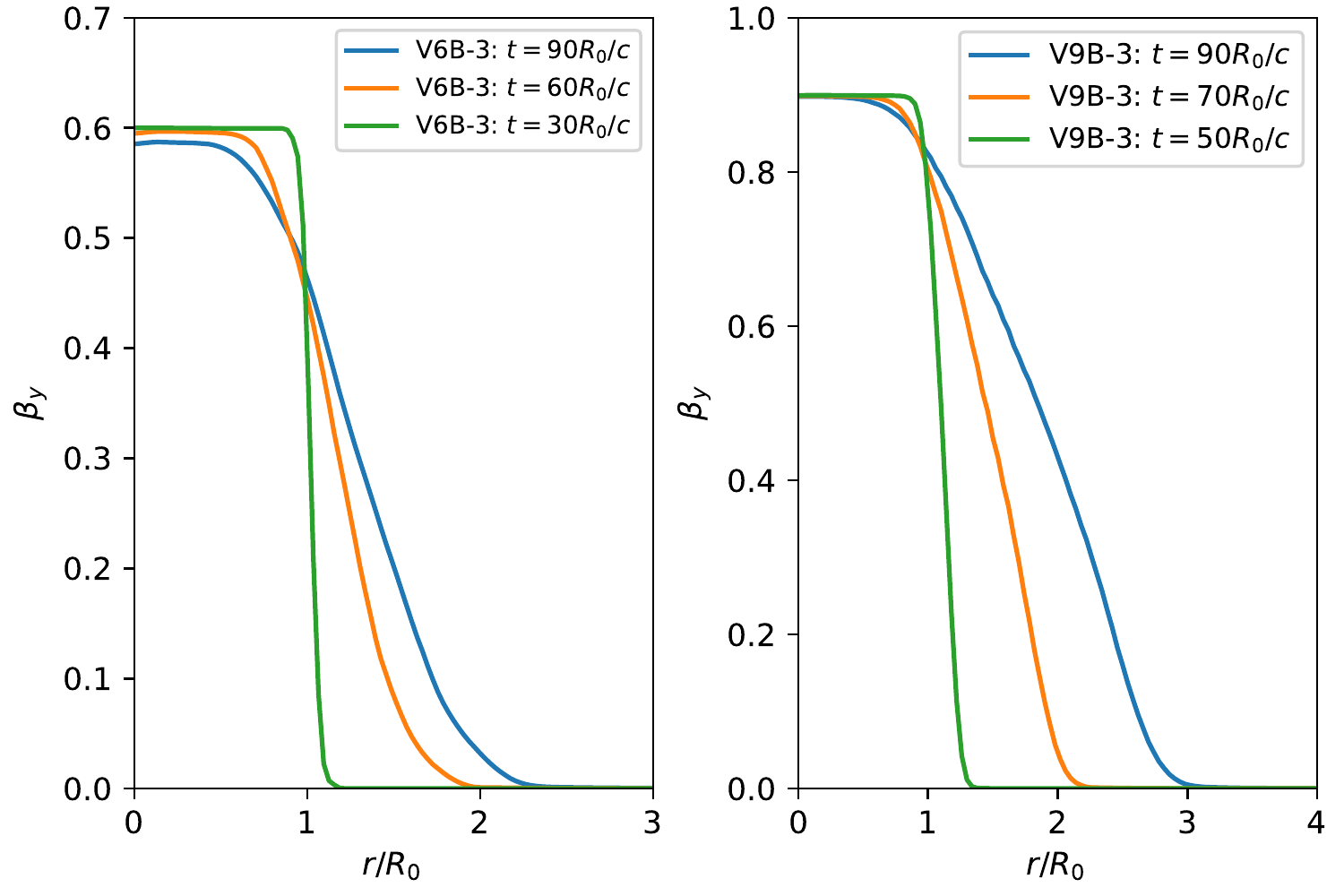}
    \caption{The evolution of the averaged axial velocity profile at different stages of development of KHI for V6B-3 case (left panel) and V9B-3 case (right panel). }
    \label{fig:V69time}
\end{figure}

We first study the evolution of velocity profiles in the linear stage, the transition stage, and the deep saturation stage of KHI. 
The choices of simulation time for these stages are based on the evolution of $E_{K,\perp}$ in Fig. \ref{fig:test} and \ref{fig:v69dy}, respectively. 
Two examples, V6B-3 and V9B-3 cases, are shown in Fig. \ref{fig:V69time}. 
In the early stage, the profile does not deviate much from the initial setup. 
The shrinking of the jet spine and the development of a sheath can be clearly observed in the transition stage, and the trend continues in the saturation stage due to the pressure imbalance induced by the saturated KHI.

\begin{figure}
    \centering
    \includegraphics[width=0.48\textwidth]{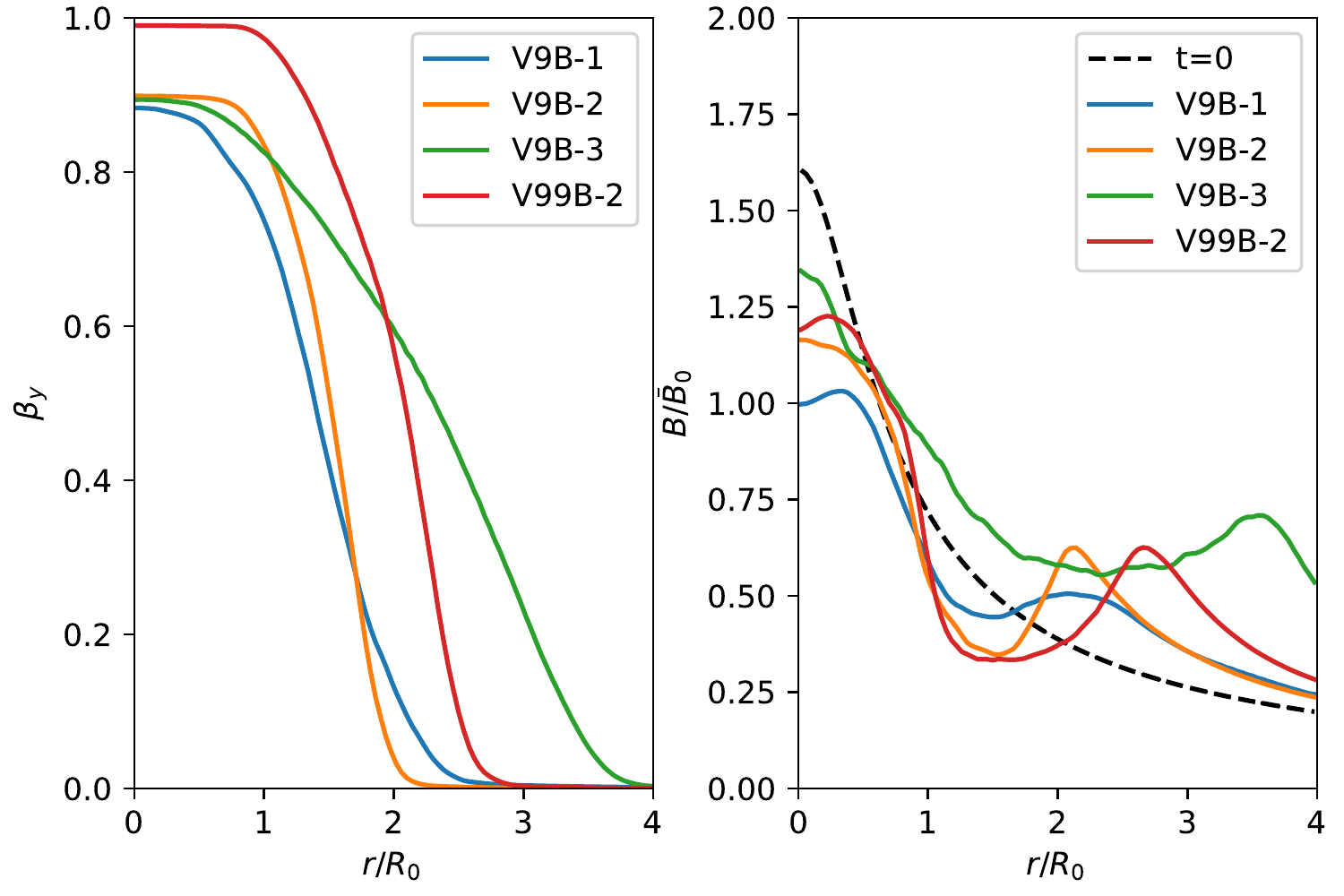}
    \caption{The axial velocity profiles (left panel) and magnitude of the magnetic field profiles (right panel) for FR II type jets at $t=90\rj/c$ (V9B-1), $t=100\rj/c$ (V9B-2), $t=120\rj/c$ (V9B-3), and $t=120\rj/c$ (V99B-2). The magnetic field is normalised to $\Bar{B}_0=\sqrt{2\sigma}B_0$.}
    \label{fig:V9B}
\end{figure}

We next focus on the jet velocity and magnetic field profiles in the deep saturation stage. 
The velocity and magnetic field profiles for FR II type jets are shown in Fig. \ref{fig:V9B}.
Our results show that the gas-pressure dominated (V9B-3) jet has the widest jet profile, while the jet widths are roughly comparable in those cases with significant contribution of magnetic pressure (V9B-1 and V9B-2). 
In the V9B-3 case, the jet sheath grows to $W_{\rm sh}\sim3.1\rj$ with a spine radius $R_{\rm sp}\sim0.6\rj$. 
In the V9B-1 and V9B-2 cases, the sheath widths are $W_{\rm sh}\sim 1.3-2.0\rj$ with spine radii $R_{\rm sp}\sim 0.5-0.8\rj$. 
The spine velocity profiles are rather flat, while the sheath exhibits a smoothly decreasing velocity profile.
Compared to the V9B-2 case, the V99B-2 case reveals a significantly wider sheath ($W_{\rm sh}\sim2.0\rj$) but has a similar spine width in the deep saturation stage.

For the magnetic field profiles, we also show the initial profile at $t=0$ in the right panel of Fig. \ref{fig:V9B}, which is identical in all cases. 
The magnetic field is reduced in the spine, while it is enhanced within and even beyond the sheath in the saturated KHI stage with a considerable pile up of magnetic field at the sheath edge in all cases.
In higher magnetization cases, the magnetic field in the spine decreases further and its pile up effect at the sheath edge is weaker. 
Besides, a global amplification of the magnetic field is evident in the gas-pressure dominated case - V9B-3 with a more uniform sheath magnetic field profile.

\begin{figure}
    \centering
    \includegraphics[width=0.48\textwidth]{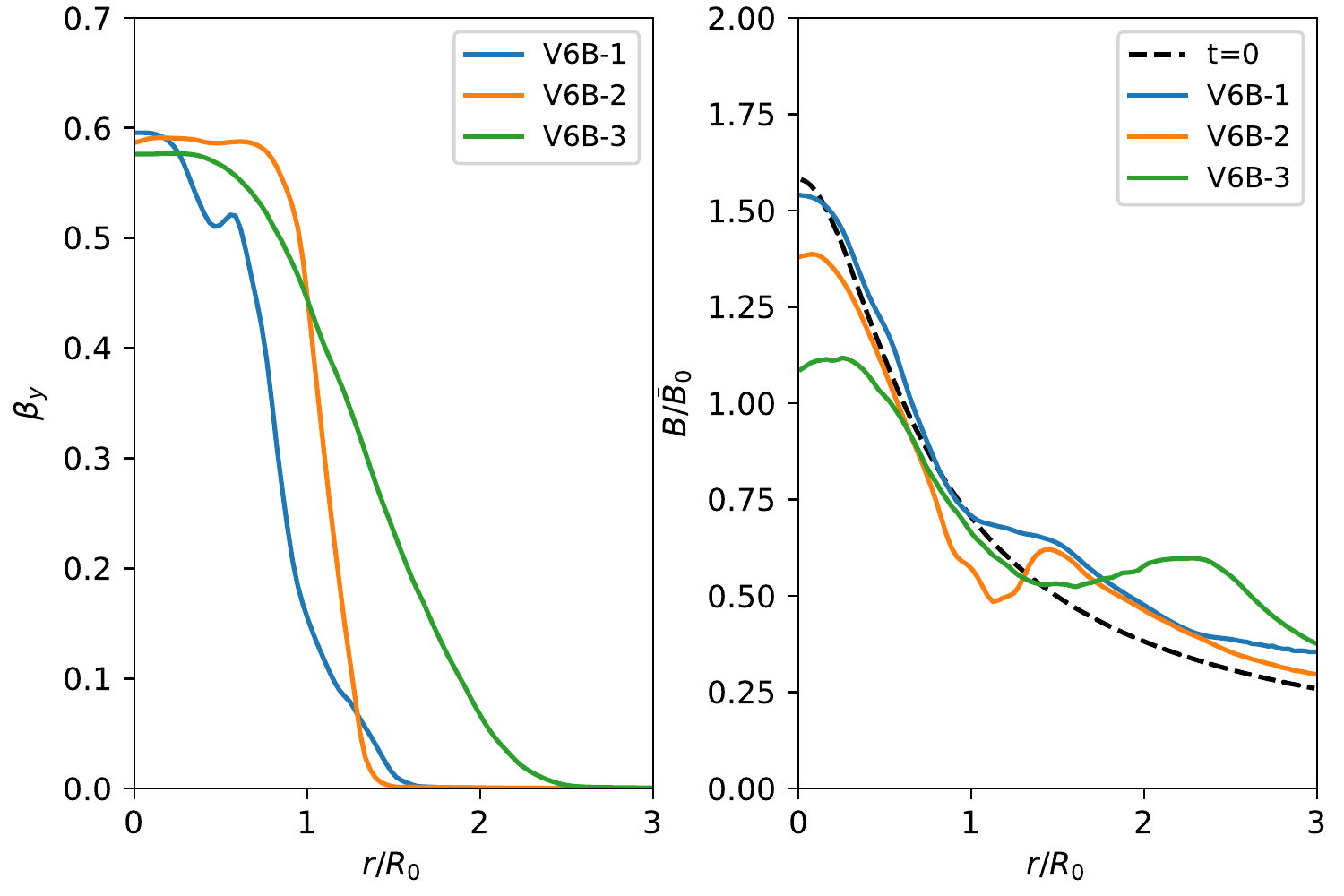}
    \caption{The velocity profiles (left panel) and magnetic field profiles (right panels) are presented for FR I type jets at $t=90\rj/c$ (V6B-1), $t=100\rj/c$ (V6B-2), $t=120\rj/c$ (V6B-3). The magnetic field is normalised with $\Bar{B}_0=\sqrt{2\sigma}B_0$. }
    \label{fig:V6B}
\end{figure}

The results for the simulated FR I jets in the deep KHI saturation stage are shown in Fig. \ref{fig:V6B}. 
As with the simulated FR II jets, the spines shrink to $R_{\rm sp}\sim0.5-0.7\rj$ with expanding sheaths of $W_{\rm sh}\sim0.7-2\rj $, and the magnetic field increases in the sheath and piles up near the sheath edge. 
In contrast to FR II jets, the magnetic field decreases more in the weaker magnetized spines. 
The velocity profiles are smoothly decreasing in the sheaths, though can be quite different in the spines. 
In the magnetic pressure dominated case - V6B-1, the spine velocity profile has an inflection point, though 
its magnetic field profile deviates only slightly from the initial configuration. 
On the other hand, in the V6B-2 and V6B-3 cases where the gas-pressure is important, the spine velocity 
profiles are flat. 

Overall, the velocity profile in the KHI-saturation stage is flat in the spine for most cases, and decreases smoothly in the sheath, with a roughly linear profile over much of its width.
In the deep saturation regime, the sheath size is wider than the spine, i.e., $W_{\rm sh}\gtrsim R_{\rm sp}$. 
In general, in the KHI-saturation stage a higher velocity leads to a wider sheath for fixed magnetisation, while for fixed velocities the gas pressure-dominated jet has the widest sheath.

\subsection{Turbulence}\label{sec:turbulence}
 
We next examine the energy spectra of the jet velocity and the magnetic field in the KHI saturation stage.
To compare with the Fokker-Planck approach of \citet{Wang2021MNRAS} we perform a fast Fourier transform (FFT) of the fluctuating velocity and magnetic fields along the jet axis with wave-number $k_\parallel = 2 \pi / \lambda$, where $\lambda$ is the corresponding wavelength. The minimum wavenumber ($k_{\parallel,\rm min}$) thus corresponds to the wavelength $\lambda$ equal to the simulation box size (see Table \ref{tab:1}).

\begin{figure}
    \centering
    \includegraphics[width=0.48\textwidth]{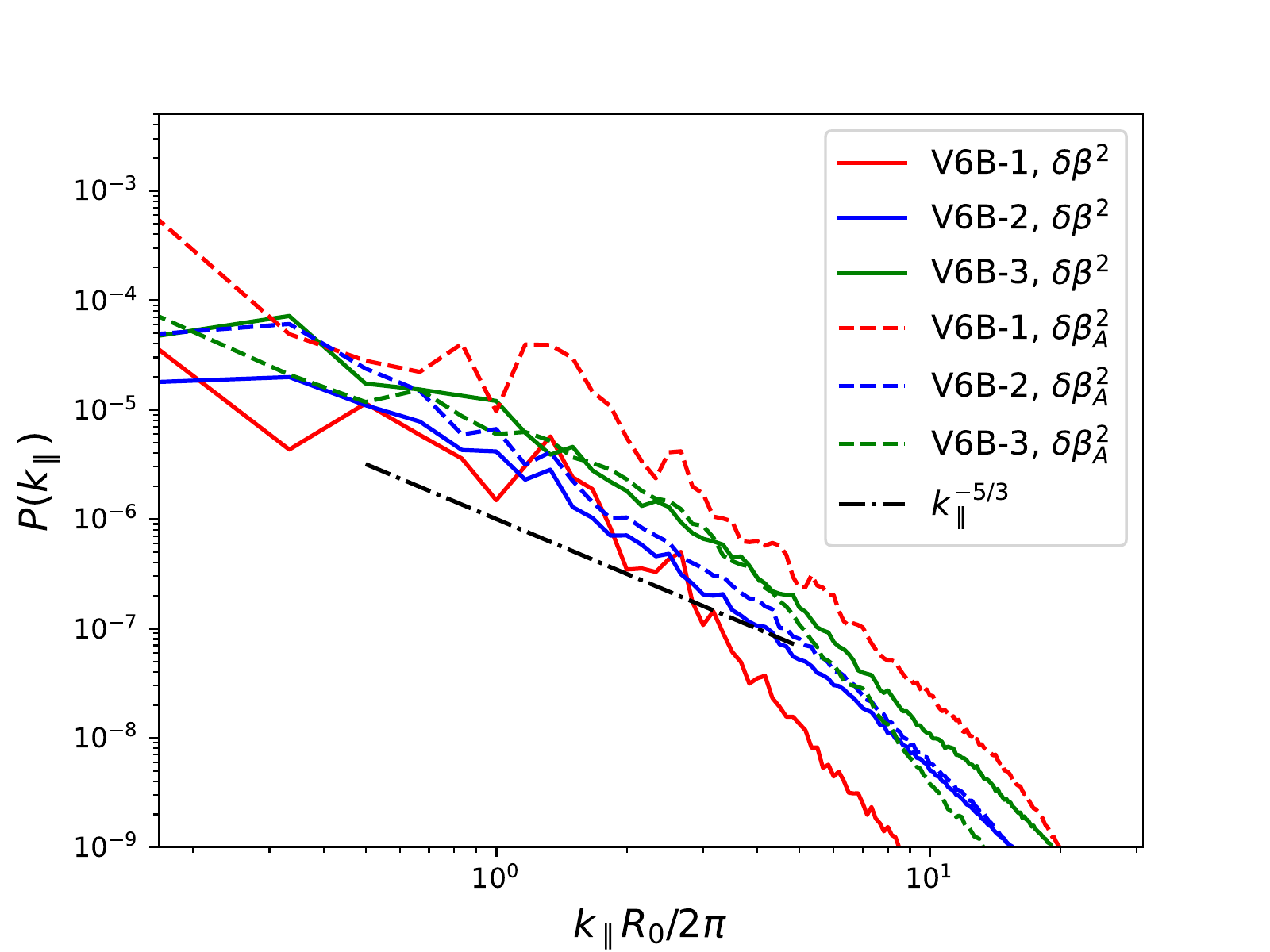}
    \includegraphics[width=0.48\textwidth]{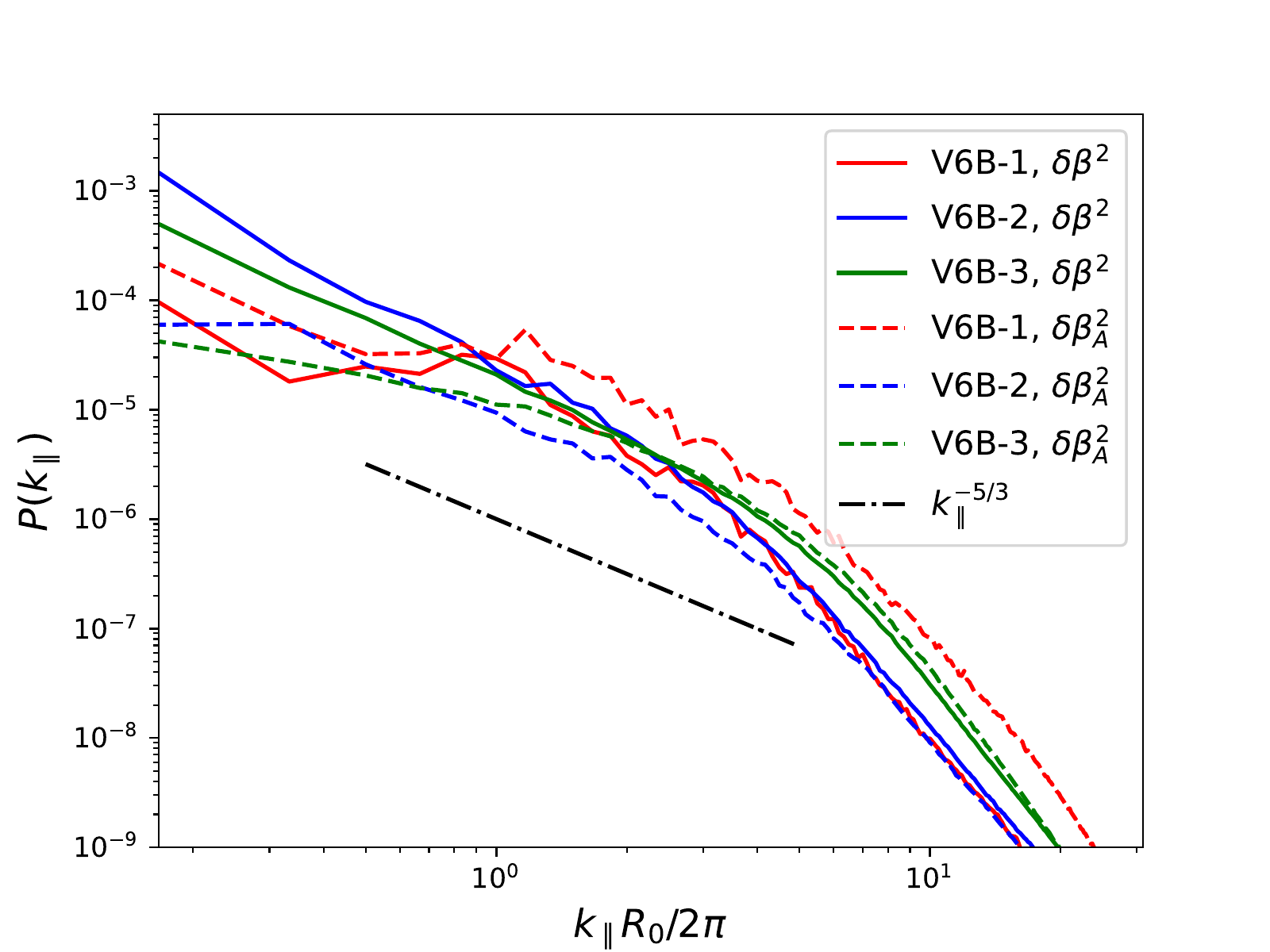}
    \caption{
    1D axial turbulence spectra of simulated FR I jet spines within $r=0.6,~0.7,~0.5 \rj$ (top panel) and their spine-sheaths in $r=1.5,~1.4~,2.5 \rj$ (bottom panel) are presented for V6B-1, V6B-2, and V6B-3 cases respectively.
    The solid lines trace the power spectrum of the turbulent velocity field $\delta\beta$, and the dashed lines show the power spectrum of the turbulent Alfv\'en velocity $\delta\beta_A$, which we take to represent the turbulent magnetic field.
    For comparison, we also show the benchmark Kolmogorov-type $k_\parallel^{-5/3}$ turbulence.}
    \label{fig:Turb-fr1}
\end{figure}

\begin{figure}
    \centering
    \includegraphics[width=0.48\textwidth]{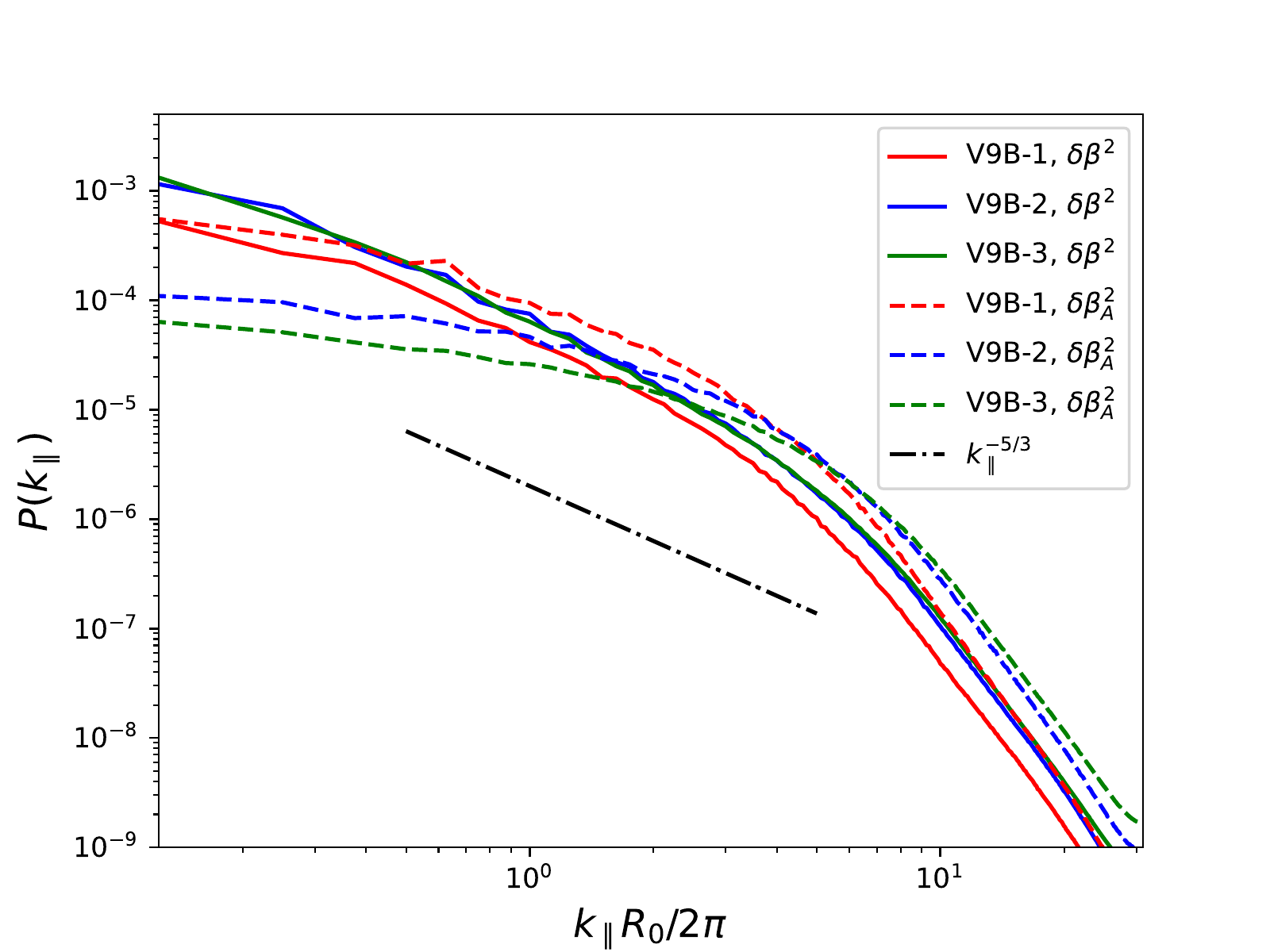}
    \includegraphics[width=0.48\textwidth]{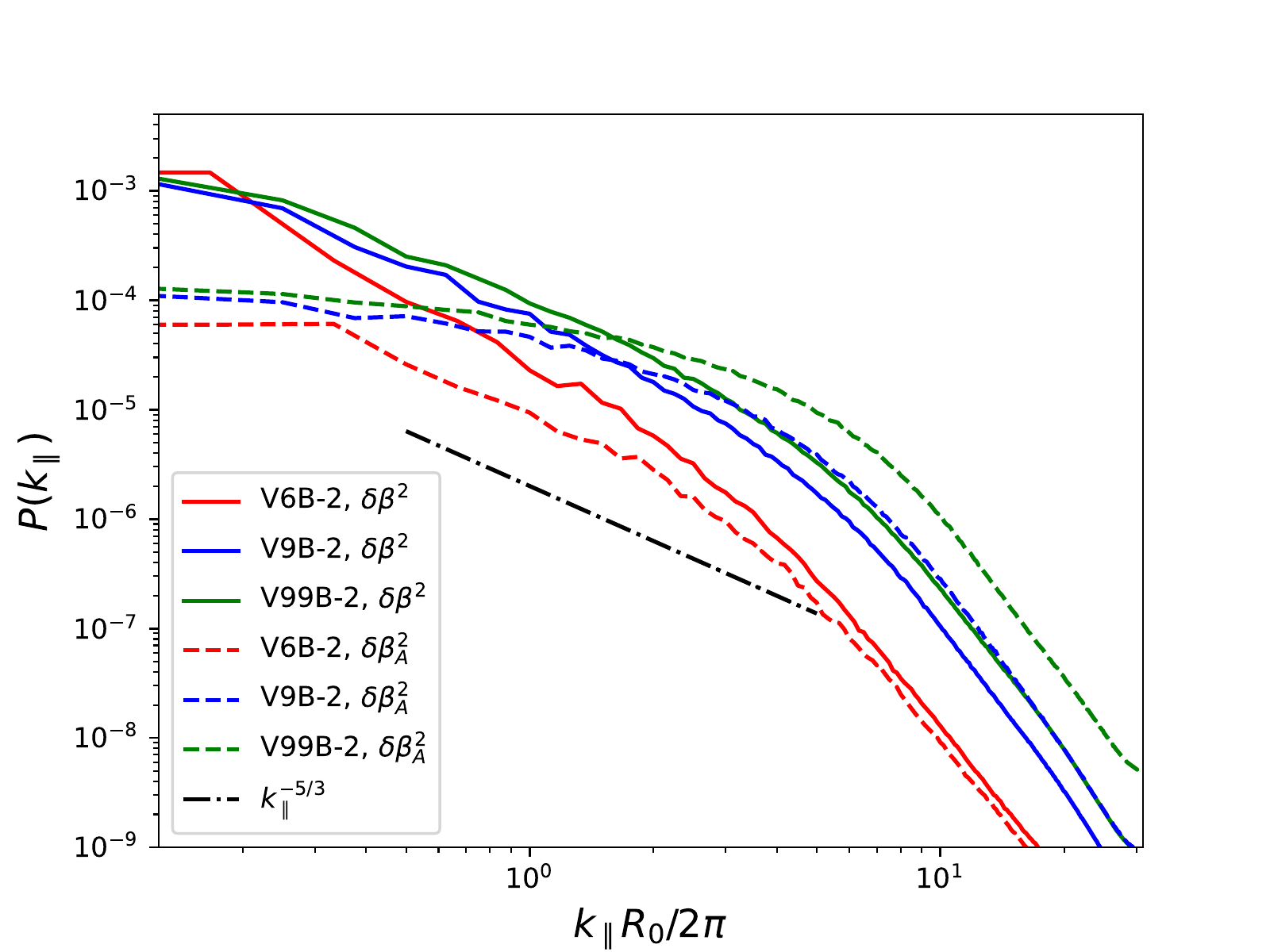}
    \caption{
    1D axial turbulent spectra of simulated FR II jets within $r=2.5,~2.1,~3.7\rj$ (top panel) for V9B-1, V9B-2, and V9B-3 cases and simulated jets with fixed magnetization parameter but different velocities (bottom panel) are presented. The jet width of V99B-2 case is chosen to be $r=2.8\rj$.
    The solid lines are for the turbulent velocity field $\delta\beta$, and the dashed lines are for the turbulent Alfv\'en velocity field $\delta\beta_A$.
    For comparison, we also show the benchmark Kolmogorov-type $k_\parallel^{-5/3}$ turbulence.}
    \label{fig:Turb-fr2}
\end{figure}

The magnetic turbulence is normalised according to the Alfv\'en velocity $\beta_A=B/\sqrt{4\pi\Gamma\rho}c$, where $B$ is the magnitude of the magnetic field in each computational cell. This allows us to make straightforward comparison with the FFT of the magnitude of the velocity field.
We consider the turbulence inside the spine ($r\leq0.7\rj$ for the simulated FR I jets) as well as in the spine-sheath region.
For the latter, we include the data within $2.5\rj$ for FR I type jets, and $3.7\rj$ for FR II type jets. 
The FFT analyses are performed for both FR I/II jets with different magnetizations. 
The results are shown in Fig. \ref{fig:Turb-fr1} and \ref{fig:Turb-fr2}. 
Due to the limited dynamic range of the simulation,  the spectrum covers only two orders of magnitude in $k_\|$. 
The roll-over at high wave numbers, $k_{\parallel}\rj/2 \pi\gtrsim 6$ is an artefact of the limited spatial resolution of the simulations.

In the range of $k_{\parallel}\rj/2 \pi\lesssim6$, the power-spectrum of the velocity field is found to be approximately consistent with Kolmogorov-type turbulence for all cases, $P(k_\|) \propto k_\|^{-5/3}$. 
The magnetic power spectra are noticeably flatter at small $k_\|$ in virtually all cases, as the KHI driven field fluctuations are concentrated on sub-jet radius scales. At larger $k_\|$, the magnetic and kinetic power spectra converge. We anticipate that with increased spatial resolution, both the magnetic and velocity field power spectra would extend to larger $k_\|$ with Kolmogorov scaling.

The amplitudes of the velocity field power spectra at a given $k_\|$ vary slightly for different magnetizations and jet velocities, with amplitudes at the minimum wave-number being $\delta \beta^2(k_{\parallel,\rm min})\sim (10^{-3}-10^{-4})$ for the spine-sheath jets as shown in Fig. \ref{fig:Turb-fr1} and \ref{fig:Turb-fr2}. 
The turbulent magnetic energy density is slightly higher than the turbulent kinetic energy density in the V6B-1 and V9B-1 cases and lower in other cases, but overall they are roughly comparable.
Fig. \ref{fig:Turb-fr1} and \ref{fig:Turb-fr2} also show that the highest magnetization cases have the lowest turbulent kinetic energy, and the magnetic turbulence is enhanced on small scales in the lowest magnetization cases.  

In the top and bottom panel of Fig. \ref{fig:Turb-fr1}, we show the turbulence spectra in the spine and spine-sheath of the simulated FR I jets respectively. 
We found that the turbulence is well developed in both the spine and sheath regions.
In the top panel of Fig. \ref{fig:Turb-fr2}, we show the turbulent spectra in the spine-sheath of FR II jets, whose behavior are similar to those of FR I jets.
The bottom panel of Fig. \ref{fig:Turb-fr2}, compares the turbulent spectra with the same magnetization but different velocities. 
This shows that the turbulent magnetic and kinetic energy densities are larger in higher velocity jets. 

\subsection{Particle spectral index of shear acceleration}
Following our previous paper \citep{Wang2021MNRAS,Rieger2021,Rieger2022ApJ}, 
the spectral index of the accelerated particle distribution, $n(\gamma)\propto \gamma^s$, 
in the leaky-box framework for gradual shear acceleration is 
\begin{equation}
    s=\frac{q-1}{2}- \sqrt{\frac{(q-5)^2}{4}+w},\label{eq.s-minus}
\end{equation}
which depends on the turbulence spectrum $k_\parallel^{-q}$, and we find $q\approx5/3$ for 
most cases in our simulations.
The shear coefficient $w$,  is determined by the region ($\Delta r$) that confines the particles, 
and the flow velocity profile $\beta_y(r)$, as
\begin{equation}
    w= \frac{10}{\Gamma^4(r)\Delta r^2 }\left(\frac{\partial 
\beta_y(r)}{\partial r}\right)^{-2},
\label{Diffu_shear}
\end{equation}
where $\Gamma(r)$ is the jet flow Lorentz factor.
To approximate spatial transport, we replace $w$ by an averaged shear coefficient ($\bar{w}$) 
for cylindrical jets, defined here as the average of $< \Gamma^2(r)\partial \beta_y(r)/\partial r>$ 
over $r$, with $<f>=\int r f dr \big/ \int r dr$. 
Other averaging methods have been explored by assuming a linear-decreasing profile \citep{Rieger2019ApJ} 
and non-linear type profiles \citep{Rieger2022ApJ} in the sheath.

We approximate our simulation results as a simple velocity profile with $\beta_y=\beta_0$ inside the spine ($r\leq R_{\rm sp}$) and $\beta_y=\beta_0 (R_{\rm jet}-r)/(W_{\rm sh})$ in the sheath with $R_{\rm jet}= R_{\rm sp}+W_{\rm sh}$ being the total jet radius. 
As particles, entering the spine, could re-enter the sheath again, and as turbulence is well developed in both the spine and the sheath, 
here we consider $\Delta r=R_{\rm jet}$. 
This then yields 
\begin{equation}
    \bar{w}=\frac{10 \beta_0^2}{[\ln\Gamma_0^{-2}W_{\rm sh}R_{\rm jet}^{-1} +2 \beta_0 \tanh ^{-1}(\beta_0)]^2}.\label{eq.wbar}
\end{equation}
For $\beta_0\rightarrow 1$, we have $w\rightarrow0$ and $s\rightarrow-4/3$. 

\begin{figure}
    \centering
    \includegraphics[width=0.48\textwidth]{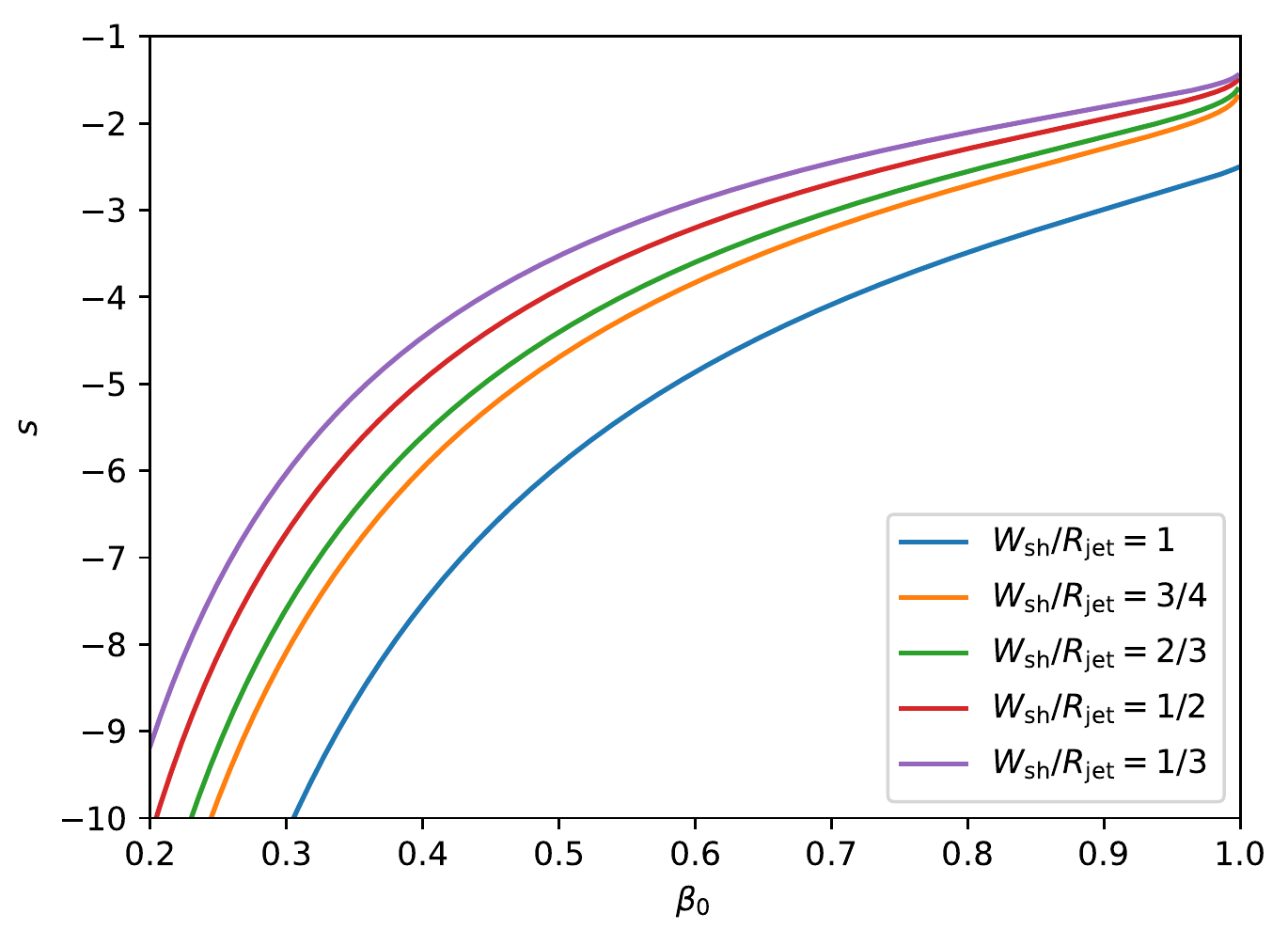}
    \caption{Spectral index of shear-accelerated particle distribution, $n(\gamma)\propto \gamma^s$, for different ratios of $W_{\rm sh}/R_{\rm jet}=1,~1/2,~1/3,~1/4$ and spine velocities.}
    \label{fig:shear-index}
\end{figure}

In Fig. \ref{fig:shear-index}, we show the resultant spectral index of the accelerated particles for different $W_{\rm sh}/R_{\rm jet}$,
where the blue line ($W_{\rm sh}=R_{\rm jet}$) is equivalent to the case that only considers the sheath. 
We find that at the same velocity, the spectrum is harder for a narrower sheath, and in the highly relativistic limit, the spectral index 
is less dependent on the velocity profile, consistent with \cite{Rieger2022ApJ}. 

From our simulation results, we find $ W_{\rm sh}/R_{\rm jet}\sim1/4-1/2$ in the transition stage and 
$W_{\rm sh}/R_{\rm jet}\sim 1/2-4/5$ in the deep saturation stage for FR I/II type jets, depending on 
the magnetization and the velocity. 
In our previous studies, estimates of the magnetic field and the coefficient $\bar{w}$ have been 
obtained for 3C 273 and Centaurus A by fitting their multi-wavelength SEDs \citep[see Table1 of][]{Wang2021MNRAS}. 
Our results here suggest that Centaurus~A and 3C~273 can be approximately described by V6B-2 case and V9B-3 case, respectively. 
Taking $W_{\rm sh}/R_{\rm jet}=1/2~(3/4)$ for Centaurus A (3C 273), $\bar{w}=15~(4.7)$ 
requires $\beta_0=0.5~(0.87)$. 
These velocities are consistent with radio observations, that indicate $\beta_0\sim0.5$ for 
Centaurus A \citep{Hardcastle2003,Snios2019} and $\beta_0<0.94$ for 3C 273 \citep{Meyer2016}.

\section{Conclusion and Discussion}\label{sec:conclusion}

Particle acceleration in shearing flows depends on the underlying flow velocity profile and turbulence 
properties. In this paper, we study such properties in self-generated spine-sheaths through RMHD 
simulations. 
By propagating a relativistic spine through a static cocoon, we observe that a sheath is
formed at the interface, mainly due to the KHI. 
We find that in general a higher velocity or a lower magnetisation leads to a wider 
sheath, and in the transition and saturation stage of the KHI, the simulated FR I/II 
jets have sheath widths $W_{\rm sh}/R_{\rm jet}\sim1/4-1/2$ and $ W_{\rm sh}/R_{\rm jet}
\sim1/2-4/5$ respectively. 
The averaged velocity is approximately constant in the spine and shows a smooth (approximately linear) decrease in the sheath.

Development of turbulence is seen in both the spine and the sheath. 
The turbulent velocity field shows a rather good agreement with Kolmogorov-type behaviour in all cases, independent of the initial velocity and magnetization. 
The magnetic field, on the other hand, reveals some dependence on magnetization and is of flatter turbulence spectrum. 
In general, we find that the turbulent magnetic energy is roughly comparable to the turbulent kinetic energy. 

The inferred flow properties have been used to explore the potential of shear particle 
acceleration in large-scale AGN jets. Assuming a quasi-linearly decreasing profile 
in the sheath, the spine velocities required to reproduce the observed X-ray spectra 
are found to be compatible with values reported for Centaurus A and 3C 273. We note
that our simulations provide indications that the velocity profiles of the spine/sheath
before averaging can be more complex, e.g. see Fig. \ref{fig:V9B-3map} and 
\ref{fig:V6B-2-y-time}.
A higher velocity gradient in some local region could enhance the 
efficiency of shear acceleration. A quantification of this is, however, beyond 
the scope of the present paper, and is left to future test-particle simulations.

From a phenomenological perspective, one of the major challenges in understanding the FR dichotomy is that most 
of the observed SEDs of FR I jets can be explained by a single population of electrons 
\citep[e.g.,][]{Perlman2001,Hardcastle2001,Sun2018}, while two populations of 
electrons seem to be required to account for the SEDs of FR II jets \citep[e.g.,][]{Jester2006,Hardcastle2006}. 
One interesting possibility is that shocks, formed during the KHI development, provide seed particles for shear acceleration to generate a second population of electrons. 
To investigate this, we also searched for shocks in our simulations following the approach in \citet{Mukherjee2020MNRAS,Mukherjee2021MNRAS}. 
We find shocks in the KHI-saturation stage primarily in the lowest-magnetisation 
cases $\sigma=10^{-3}$. 
Since, as our results indicate, Centaurus~A and 3C 273 may be approximately described  
by the V6B-2 and the V9B-3 case, respectively, this would suggest that a second population of electrons should only be present in 3C 273, which would be consistent with observation. 
In our simulations, however, these shocks seem to occupy only a very small volume compared with the whole simulation domain. 
Since this outcome might be affected by our choice of periodic boundary conditions,
further high-resolution, full jet propagation studies will be needed to adequately 
conclude on this possibility. 
This will also help to better understand the formation of knots and the possibility of 
variable X-ray emission in large-scale AGN jets.

\section*{Acknowledgements}
We thank the referee for helpful suggestions and Eileen Meyer, Zhi Li, Omer Bromberg, and Krzysztof Nalewajko for helpful discussions.
JSW acknowledges the support from the Alexander von Humboldt Foundation. 
YM acknowledges the support by the National Natural Science Foundation of China (Grant No. 12273022).
FMR acknowledges support by a DFG Fellowship (RI 1187/8-1).

\section*{Data Availability}
The data underlying this article will be shared on reasonable request to the corresponding author.

\label{lastpage}

\bibliographystyle{mnras}
\bibliography{ref}

\begin{thebibliography}{}
\makeatletter
\relax
\def\mn@urlcharsother{\let\do\@makeother \do\$\do\&\do\#\do\^\do\_\do\%\do\~}
\def\mn@doi{\begingroup\mn@urlcharsother \@ifnextchar [ {\mn@doi@}
  {\mn@doi@[]}}
\def\mn@doi@[#1]#2{\def\@tempa{#1}\ifx\@tempa\@empty \href
  {http://dx.doi.org/#2} {doi:#2}\else \href {http://dx.doi.org/#2} {#1}\fi
  \endgroup}
\def\mn@eprint#1#2{\mn@eprint@#1:#2::\@nil}
\def\mn@eprint@arXiv#1{\href {http://arxiv.org/abs/#1} {{\tt arXiv:#1}}}
\def\mn@eprint@dblp#1{\href {http://dblp.uni-trier.de/rec/bibtex/#1.xml}
  {dblp:#1}}
\def\mn@eprint@#1:#2:#3:#4\@nil{\def\@tempa {#1}\def\@tempb {#2}\def\@tempc
  {#3}\ifx \@tempc \@empty \let \@tempc \@tempb \let \@tempb \@tempa \fi \ifx
  \@tempb \@empty \def\@tempb {arXiv}\fi \@ifundefined
  {mn@eprint@\@tempb}{\@tempb:\@tempc}{\expandafter \expandafter \csname
  mn@eprint@\@tempb\endcsname \expandafter{\@tempc}}}

\bibitem[\protect\citeauthoryear{{Baty} \& {Keppens}}{{Baty} \&
  {Keppens}}{2002}]{Baty2002ApJ}
{Baty} H.,  {Keppens} R.,  2002, \mn@doi [\apj] {10.1086/343893}, \href
  {https://ui.adsabs.harvard.edu/abs/2002ApJ...580..800B} {580, 800}

\bibitem[\protect\citeauthoryear{{Berezhko}}{{Berezhko}}{1981}]{Berezhko1981}
{Berezhko} E.~G.,  1981, ZhETF Pisma Redaktsiiu, \href
  {https://ui.adsabs.harvard.edu/abs/1981ZhPmR..33..416B} {33, 416}

\bibitem[\protect\citeauthoryear{{Berezhko} \& {Krymskii}}{{Berezhko} \&
  {Krymskii}}{1981}]{BerezhkoKrymskii1981}
{Berezhko} E.~G.,  {Krymskii} G.~F.,  1981, Soviet Astronomy Letters, \href
  {https://ui.adsabs.harvard.edu/abs/1981SvAL....7..352B} {7, 352}

\bibitem[\protect\citeauthoryear{{Birkinshaw}}{{Birkinshaw}}{1991}]{Birkinshaw1991MNRAS}
{Birkinshaw} M.,  1991, \mn@doi [\mnras] {10.1093/mnras/252.4.505}, \href
  {https://ui.adsabs.harvard.edu/abs/1991MNRAS.252..505B} {252, 505}

\bibitem[\protect\citeauthoryear{{Blandford}, {Meier}  \&
  {Readhead}}{{Blandford} et~al.}{2019}]{Blandford2019}
{Blandford} R.,  {Meier} D.,   {Readhead} A.,  2019, \mn@doi [\araa]
  {10.1146/annurev-astro-081817-051948}, \href
  {https://ui.adsabs.harvard.edu/abs/2019ARA&A..57..467B} {57, 467}

\bibitem[\protect\citeauthoryear{{Boccardi}, {Krichbaum}, {Bach}, {Mertens},
  {Ros}, {Alef}  \& {Zensus}}{{Boccardi} et~al.}{2016}]{Boccardi2016}
{Boccardi} B.,  {Krichbaum} T.~P.,  {Bach} U.,  {Mertens} F.,  {Ros} E.,
  {Alef} W.,   {Zensus} J.~A.,  2016, \mn@doi [\aap]
  {10.1051/0004-6361/201526985}, \href
  {https://ui.adsabs.harvard.edu/abs/2016A&A...585A..33B} {585, A33}

\bibitem[\protect\citeauthoryear{{Bodo}, {Mamatsashvili}, {Rossi}  \&
  {Mignone}}{{Bodo} et~al.}{2013}]{Bodo2013MNRAS}
{Bodo} G.,  {Mamatsashvili} G.,  {Rossi} P.,   {Mignone} A.,  2013, \mn@doi
  [\mnras] {10.1093/mnras/stt1225}, \href
  {https://ui.adsabs.harvard.edu/abs/2013MNRAS.434.3030B} {434, 3030}

\bibitem[\protect\citeauthoryear{{Bodo}, {Mamatsashvili}, {Rossi}  \&
  {Mignone}}{{Bodo} et~al.}{2019}]{Bodo2019MNRAS}
{Bodo} G.,  {Mamatsashvili} G.,  {Rossi} P.,   {Mignone} A.,  2019, \mn@doi
  [\mnras] {10.1093/mnras/stz591}, \href
  {https://ui.adsabs.harvard.edu/abs/2019MNRAS.485.2909B} {485, 2909}

\bibitem[\protect\citeauthoryear{{Borse}, {Acharya}, {Vaidya}, {Mukherjee},
  {Bodo}, {Rossi}  \& {Mignone}}{{Borse} et~al.}{2021}]{Borse2021A&A}
{Borse} N.,  {Acharya} S.,  {Vaidya} B.,  {Mukherjee} D.,  {Bodo} G.,  {Rossi}
  P.,   {Mignone} A.,  2021, \mn@doi [\aap] {10.1051/0004-6361/202140440},
  \href {https://ui.adsabs.harvard.edu/abs/2021A&A...649A.150B} {649, A150}

\bibitem[\protect\citeauthoryear{{Breiding}, {Meyer}, {Georganopoulos},
  {Keenan}, {DeNigris}  \& {Hewitt}}{{Breiding} et~al.}{2017}]{Breiding2017}
{Breiding} P.,  {Meyer} E.~T.,  {Georganopoulos} M.,  {Keenan} M.~E.,
  {DeNigris} N.~S.,   {Hewitt} J.,  2017, \mn@doi [\apj]
  {10.3847/1538-4357/aa907a}, \href
  {https://ui.adsabs.harvard.edu/abs/2017ApJ...849...95B} {849, 95}

\bibitem[\protect\citeauthoryear{{Bromberg}, {Singh}, {Davelaar}  \&
  {Philippov}}{{Bromberg} et~al.}{2019}]{Bromberg2019ApJ}
{Bromberg} O.,  {Singh} C.~B.,  {Davelaar} J.,   {Philippov} A.~A.,  2019,
  \mn@doi [\apj] {10.3847/1538-4357/ab3fa5}, \href
  {https://ui.adsabs.harvard.edu/abs/2019ApJ...884...39B} {884, 39}

\bibitem[\protect\citeauthoryear{{Cara} et~al.,}{{Cara}
  et~al.}{2013}]{Cara2013}
{Cara} M.,  et~al., 2013, \mn@doi [\apj] {10.1088/0004-637X/773/2/186}, \href
  {https://ui.adsabs.harvard.edu/abs/2013ApJ...773..186C} {773, 186}

\bibitem[\protect\citeauthoryear{{Chow}, {Davelaar}  \& {Sironi}}{{Chow}
  et~al.}{2022}]{Chow2022arXiv}
{Chow} A.,  {Davelaar} J.,   {Sironi} L.,  2022, arXiv e-prints, \href
  {https://ui.adsabs.harvard.edu/abs/2022arXiv220913699C} {p. arXiv:2209.13699}

\bibitem[\protect\citeauthoryear{{Earl}, {Jokipii}  \& {Morfill}}{{Earl}
  et~al.}{1988}]{Earl1988}
{Earl} J.~A.,  {Jokipii} J.~R.,   {Morfill} G.,  1988, \mn@doi [\apjl]
  {10.1086/185242}, \href
  {https://ui.adsabs.harvard.edu/abs/1988ApJ...331L..91E} {331, L91}

\bibitem[\protect\citeauthoryear{{Ferrari}, {Trussoni}  \&
  {Zaninetti}}{{Ferrari} et~al.}{1978}]{Ferrari1978A&A}
{Ferrari} A.,  {Trussoni} E.,   {Zaninetti} L.,  1978, \aap, \href
  {https://ui.adsabs.harvard.edu/abs/1978A&A....64...43F} {64, 43}

\bibitem[\protect\citeauthoryear{{Gabuzda}, {Reichstein}  \&
  {O'Neill}}{{Gabuzda} et~al.}{2014}]{Gabuzda2014}
{Gabuzda} D.~C.,  {Reichstein} A.~R.,   {O'Neill} E.~L.,  2014, \mn@doi
  [\mnras] {10.1093/mnras/stu1381}, \href
  {https://ui.adsabs.harvard.edu/abs/2014MNRAS.444..172G} {444, 172}

\bibitem[\protect\citeauthoryear{{Georganopoulos}, {Meyer}  \&
  {Perlman}}{{Georganopoulos} et~al.}{2016}]{Georganopoulos2016}
{Georganopoulos} M.,  {Meyer} E.,   {Perlman} E.,  2016, \mn@doi [Galaxies]
  {10.3390/galaxies4040065}, \href
  {https://ui.adsabs.harvard.edu/abs/2016Galax...4...65G} {4, 65}

\bibitem[\protect\citeauthoryear{{Gourgouliatos} \&
  {Komissarov}}{{Gourgouliatos} \& {Komissarov}}{2018}]{2018NatAs...2..167G}
{Gourgouliatos} K.~N.,  {Komissarov} S.~S.,  2018, \mn@doi [Nature Astronomy]
  {10.1038/s41550-017-0338-3}, \href
  {https://ui.adsabs.harvard.edu/abs/2018NatAs...2..167G} {2, 167}

\bibitem[\protect\citeauthoryear{{H.~E.~S.~S. Collaboration}
  et~al.,}{{H.~E.~S.~S. Collaboration} et~al.}{2020}]{HESS2020}
{H.~E.~S.~S. Collaboration} et~al., 2020, \mn@doi [\nat]
  {10.1038/s41586-020-2354-1}, \href
  {https://ui.adsabs.harvard.edu/abs/2020Natur.582..356H} {582, 356}

\bibitem[\protect\citeauthoryear{{Hardcastle}, {Birkinshaw}  \&
  {Worrall}}{{Hardcastle} et~al.}{2001}]{Hardcastle2001}
{Hardcastle} M.~J.,  {Birkinshaw} M.,   {Worrall} D.~M.,  2001, \mn@doi
  [\mnras] {10.1111/j.1365-2966.2001.04699.x}, \href
  {https://ui.adsabs.harvard.edu/abs/2001MNRAS.326.1499H} {326, 1499}

\bibitem[\protect\citeauthoryear{{Hardcastle}, {Worrall}, {Kraft}, {Forman},
  {Jones}  \& {Murray}}{{Hardcastle} et~al.}{2003}]{Hardcastle2003}
{Hardcastle} M.~J.,  {Worrall} D.~M.,  {Kraft} R.~P.,  {Forman} W.~R.,  {Jones}
  C.,   {Murray} S.~S.,  2003, \mn@doi [\apj] {10.1086/376519}, \href
  {https://ui.adsabs.harvard.edu/abs/2003ApJ...593..169H} {593, 169}

\bibitem[\protect\citeauthoryear{{Hardcastle}, {Kraft}  \&
  {Worrall}}{{Hardcastle} et~al.}{2006}]{Hardcastle2006}
{Hardcastle} M.~J.,  {Kraft} R.~P.,   {Worrall} D.~M.,  2006, \mn@doi [\mnras]
  {10.1111/j.1745-3933.2006.00146.x}, \href
  {https://ui.adsabs.harvard.edu/abs/2006MNRAS.368L..15H} {368, L15}

\bibitem[\protect\citeauthoryear{{Hardee}}{{Hardee}}{1979}]{Hardee1979ApJ}
{Hardee} P.~E.,  1979, \mn@doi [\apj] {10.1086/157471}, \href
  {https://ui.adsabs.harvard.edu/abs/1979ApJ...234...47H} {234, 47}

\bibitem[\protect\citeauthoryear{{Hardee}}{{Hardee}}{2007}]{Hardee2007ApJ}
{Hardee} P.~E.,  2007, \mn@doi [\apj] {10.1086/518409}, \href
  {https://ui.adsabs.harvard.edu/abs/2007ApJ...664...26H} {664, 26}

\bibitem[\protect\citeauthoryear{{Harris} \& {Krawczynski}}{{Harris} \&
  {Krawczynski}}{2006}]{Harris2006}
{Harris} D.~E.,  {Krawczynski} H.,  2006, \mn@doi [\araa]
  {10.1146/annurev.astro.44.051905.092446}, \href
  {https://ui.adsabs.harvard.edu/abs/2006ARA&A..44..463H} {44, 463}

\bibitem[\protect\citeauthoryear{{Jester}, {Harris}, {Marshall}  \&
  {Meisenheimer}}{{Jester} et~al.}{2006}]{Jester2006}
{Jester} S.,  {Harris} D.~E.,  {Marshall} H.~L.,   {Meisenheimer} K.,  2006,
  \mn@doi [\apj] {10.1086/505962}, \href
  {https://ui.adsabs.harvard.edu/abs/2006ApJ...648..900J} {648, 900}

\bibitem[\protect\citeauthoryear{{Kim}, {Balsara}, {Lyutikov}  \&
  {Komissarov}}{{Kim} et~al.}{2018}]{Kim2018}
{Kim} J.,  {Balsara} D.~S.,  {Lyutikov} M.,   {Komissarov} S.~S.,  2018,
  \mn@doi [\mnras] {10.1093/mnras/stx3065}, \href
  {https://ui.adsabs.harvard.edu/abs/2018MNRAS.474.3954K} {474, 3954}

\bibitem[\protect\citeauthoryear{{Komissarov}}{{Komissarov}}{1999}]{Komissarov1999MNRAS}
{Komissarov} S.~S.,  1999, \mn@doi [\mnras] {10.1046/j.1365-8711.1999.02783.x},
  \href {https://ui.adsabs.harvard.edu/abs/1999MNRAS.308.1069K} {308, 1069}

\bibitem[\protect\citeauthoryear{{Laing} \& {Bridle}}{{Laing} \&
  {Bridle}}{2014}]{Laing2014}
{Laing} R.~A.,  {Bridle} A.~H.,  2014, \mn@doi [\mnras]
  {10.1093/mnras/stt2138}, \href
  {https://ui.adsabs.harvard.edu/abs/2014MNRAS.437.3405L} {437, 3405}

\bibitem[\protect\citeauthoryear{{Liu}, {Rieger}  \& {Aharonian}}{{Liu}
  et~al.}{2017}]{Liu2017}
{Liu} R.-Y.,  {Rieger} F.~M.,   {Aharonian} F.~A.,  2017, \mn@doi [\apj]
  {10.3847/1538-4357/aa7410}, \href
  {https://ui.adsabs.harvard.edu/abs/2017ApJ...842...39L} {842, 39}

\bibitem[\protect\citeauthoryear{{Lyubarskii}}{{Lyubarskii}}{1999}]{Lyubarskii1999MNRAS}
{Lyubarskii} Y.~E.,  1999, \mn@doi [\mnras] {10.1046/j.1365-8711.1999.02763.x},
  \href {https://ui.adsabs.harvard.edu/abs/1999MNRAS.308.1006L} {308, 1006}

\bibitem[\protect\citeauthoryear{{Mart{\'\i}}}{{Mart{\'\i}}}{2019}]{Marti2019Galax}
{Mart{\'\i}} J.-M.,  2019, \mn@doi [Galaxies] {10.3390/galaxies7010024}, \href
  {https://ui.adsabs.harvard.edu/abs/2019Galax...7...24M} {7, 24}

\bibitem[\protect\citeauthoryear{{Mart{\'\i}}, {M{\"u}ller}, {Font},
  {Ib{\'a}{\~n}ez}  \& {Marquina}}{{Mart{\'\i}} et~al.}{1997}]{Marti1997ApJ}
{Mart{\'\i}} J.~M.,  {M{\"u}ller} E.,  {Font} J.~A.,  {Ib{\'a}{\~n}ez}
  J.~M.~Z.,   {Marquina} A.,  1997, \mn@doi [\apj] {10.1086/303842}, \href
  {https://ui.adsabs.harvard.edu/abs/1997ApJ...479..151M} {479, 151}

\bibitem[\protect\citeauthoryear{{Mathews}}{{Mathews}}{1971}]{Mathews1971ApJ}
{Mathews} W.~G.,  1971, \mn@doi [\apj] {10.1086/150883}, \href
  {https://ui.adsabs.harvard.edu/abs/1971ApJ...165..147M} {165, 147}

\bibitem[\protect\citeauthoryear{{Meyer} \& {Georganopoulos}}{{Meyer} \&
  {Georganopoulos}}{2014}]{Meyer2014}
{Meyer} E.~T.,  {Georganopoulos} M.,  2014, \mn@doi [\apjl]
  {10.1088/2041-8205/780/2/L27}, \href
  {https://ui.adsabs.harvard.edu/abs/2014ApJ...780L..27M} {780, L27}

\bibitem[\protect\citeauthoryear{{Meyer}, {Georganopoulos}, {Sparks},
  {Godfrey}, {Lovell}  \& {Perlman}}{{Meyer} et~al.}{2015}]{Meyer2015}
{Meyer} E.~T.,  {Georganopoulos} M.,  {Sparks} W.~B.,  {Godfrey} L.,  {Lovell}
  J. E.~J.,   {Perlman} E.,  2015, \mn@doi [\apj]
  {10.1088/0004-637X/805/2/154}, \href
  {https://ui.adsabs.harvard.edu/abs/2015ApJ...805..154M} {805, 154}

\bibitem[\protect\citeauthoryear{{Meyer} et~al.,}{{Meyer}
  et~al.}{2016}]{Meyer2016}
{Meyer} E.~T.,  et~al., 2016, \mn@doi [\apj] {10.3847/0004-637X/818/2/195},
  \href {https://ui.adsabs.harvard.edu/abs/2016ApJ...818..195M} {818, 195}

\bibitem[\protect\citeauthoryear{{Mignone}, {Plewa}  \& {Bodo}}{{Mignone}
  et~al.}{2005}]{Mignone2005ApJS}
{Mignone} A.,  {Plewa} T.,   {Bodo} G.,  2005, \mn@doi [\apjs]
  {10.1086/430905}, \href
  {https://ui.adsabs.harvard.edu/abs/2005ApJS..160..199M} {160, 199}

\bibitem[\protect\citeauthoryear{{Mignone}, {Bodo}, {Massaglia}, {Matsakos},
  {Tesileanu}, {Zanni}  \& {Ferrari}}{{Mignone} et~al.}{2007}]{Mignone2007ApJS}
{Mignone} A.,  {Bodo} G.,  {Massaglia} S.,  {Matsakos} T.,  {Tesileanu} O.,
  {Zanni} C.,   {Ferrari} A.,  2007, \mn@doi [\apjs] {10.1086/513316}, \href
  {https://ui.adsabs.harvard.edu/abs/2007ApJS..170..228M} {170, 228}

\bibitem[\protect\citeauthoryear{{Mignone}, {Rossi}, {Bodo}, {Ferrari}  \&
  {Massaglia}}{{Mignone} et~al.}{2010}]{Mignone2010MNRAS}
{Mignone} A.,  {Rossi} P.,  {Bodo} G.,  {Ferrari} A.,   {Massaglia} S.,  2010,
  \mn@doi [\mnras] {10.1111/j.1365-2966.2009.15642.x}, \href
  {https://ui.adsabs.harvard.edu/abs/2010MNRAS.402....7M} {402, 7}

\bibitem[\protect\citeauthoryear{{Mizuno}, {Hardee}  \& {Nishikawa}}{{Mizuno}
  et~al.}{2007}]{Mizuno2007ApJ}
{Mizuno} Y.,  {Hardee} P.,   {Nishikawa} K.-I.,  2007, \mn@doi [\apj]
  {10.1086/518106}, \href
  {https://ui.adsabs.harvard.edu/abs/2007ApJ...662..835M} {662, 835}

\bibitem[\protect\citeauthoryear{{Mizuno}, {Lyubarsky}, {Nishikawa}  \&
  {Hardee}}{{Mizuno} et~al.}{2009}]{Mizuno2009ApJ}
{Mizuno} Y.,  {Lyubarsky} Y.,  {Nishikawa} K.-I.,   {Hardee} P.~E.,  2009,
  \mn@doi [\apj] {10.1088/0004-637X/700/1/684}, \href
  {https://ui.adsabs.harvard.edu/abs/2009ApJ...700..684M} {700, 684}

\bibitem[\protect\citeauthoryear{{Mizuno}, {Lyubarsky}, {Nishikawa}  \&
  {Hardee}}{{Mizuno} et~al.}{2012}]{Mizuno2012ApJ}
{Mizuno} Y.,  {Lyubarsky} Y.,  {Nishikawa} K.-I.,   {Hardee} P.~E.,  2012,
  \mn@doi [\apj] {10.1088/0004-637X/757/1/16}, \href
  {https://ui.adsabs.harvard.edu/abs/2012ApJ...757...16M} {757, 16}

\bibitem[\protect\citeauthoryear{{Mizuno}, {Hardee}  \& {Nishikawa}}{{Mizuno}
  et~al.}{2014}]{Mizuno2014ApJ}
{Mizuno} Y.,  {Hardee} P.~E.,   {Nishikawa} K.-I.,  2014, \mn@doi [\apj]
  {10.1088/0004-637X/784/2/167}, \href
  {https://ui.adsabs.harvard.edu/abs/2014ApJ...784..167M} {784, 167}

\bibitem[\protect\citeauthoryear{{Mukherjee}, {Bodo}, {Mignone}, {Rossi}  \&
  {Vaidya}}{{Mukherjee} et~al.}{2020}]{Mukherjee2020MNRAS}
{Mukherjee} D.,  {Bodo} G.,  {Mignone} A.,  {Rossi} P.,   {Vaidya} B.,  2020,
  \mn@doi [\mnras] {10.1093/mnras/staa2934}, \href
  {https://ui.adsabs.harvard.edu/abs/2020MNRAS.499..681M} {499, 681}

\bibitem[\protect\citeauthoryear{{Mukherjee}, {Bodo}, {Rossi}, {Mignone}  \&
  {Vaidya}}{{Mukherjee} et~al.}{2021}]{Mukherjee2021MNRAS}
{Mukherjee} D.,  {Bodo} G.,  {Rossi} P.,  {Mignone} A.,   {Vaidya} B.,  2021,
  \mn@doi [\mnras] {10.1093/mnras/stab1327}, \href
  {https://ui.adsabs.harvard.edu/abs/2021MNRAS.505.2267M} {505, 2267}

\bibitem[\protect\citeauthoryear{{Nagai} et~al.,}{{Nagai}
  et~al.}{2014}]{Nagai2014}
{Nagai} H.,  et~al., 2014, \mn@doi [\apj] {10.1088/0004-637X/785/1/53}, \href
  {https://ui.adsabs.harvard.edu/abs/2014ApJ...785...53N} {785, 53}

\bibitem[\protect\citeauthoryear{{Nakamura}, {Li}  \& {Li}}{{Nakamura}
  et~al.}{2007}]{Nakamura2007ApJ}
{Nakamura} M.,  {Li} H.,   {Li} S.,  2007, \mn@doi [\apj] {10.1086/510361},
  \href {https://ui.adsabs.harvard.edu/abs/2007ApJ...656..721N} {656, 721}

\bibitem[\protect\citeauthoryear{{Ortu{\~n}o-Mac{\'\i}as}, {Nalewajko},
  {Uzdensky}, {Begelman}, {Werner}, {Chen}  \&
  {Mishra}}{{Ortu{\~n}o-Mac{\'\i}as} et~al.}{2022}]{Ortuno-Macias2022ApJ}
{Ortu{\~n}o-Mac{\'\i}as} J.,  {Nalewajko} K.,  {Uzdensky} D.~A.,  {Begelman}
  M.~C.,  {Werner} G.~R.,  {Chen} A.~Y.,   {Mishra} B.,  2022, \mn@doi [\apj]
  {10.3847/1538-4357/ac6acd}, \href
  {https://ui.adsabs.harvard.edu/abs/2022ApJ...931..137O} {931, 137}

\bibitem[\protect\citeauthoryear{{Perlman}, {Biretta}, {Sparks}, {Macchetto}
  \& {Leahy}}{{Perlman} et~al.}{2001}]{Perlman2001}
{Perlman} E.~S.,  {Biretta} J.~A.,  {Sparks} W.~B.,  {Macchetto} F.~D.,
  {Leahy} J.~P.,  2001, \mn@doi [\apj] {10.1086/320052}, \href
  {https://ui.adsabs.harvard.edu/abs/2001ApJ...551..206P} {551, 206}

\bibitem[\protect\citeauthoryear{{Perlman}, {Clautice}, {Avachat}, {Cara},
  {Sparks}, {Georganopoulos}  \& {Meyer}}{{Perlman} et~al.}{2020}]{Perlman2020}
{Perlman} E.~S.,  {Clautice} D.,  {Avachat} S.,  {Cara} M.,  {Sparks} W.~B.,
  {Georganopoulos} M.,   {Meyer} E.,  2020, \mn@doi [Galaxies]
  {10.3390/galaxies8040071}, \href
  {https://ui.adsabs.harvard.edu/abs/2020Galax...8...71P} {8, 71}

\bibitem[\protect\citeauthoryear{{Perucho} \& {Mart{\'\i}}}{{Perucho} \&
  {Mart{\'\i}}}{2007}]{Perucho2007MNRAS}
{Perucho} M.,  {Mart{\'\i}} J.~M.,  2007, \mn@doi [\mnras]
  {10.1111/j.1365-2966.2007.12454.x}, \href
  {https://ui.adsabs.harvard.edu/abs/2007MNRAS.382..526P} {382, 526}

\bibitem[\protect\citeauthoryear{{Perucho}, {Hanasz}, {Mart{\'\i}}  \&
  {Sol}}{{Perucho} et~al.}{2004}]{Perucho2004A&A}
{Perucho} M.,  {Hanasz} M.,  {Mart{\'\i}} J.~M.,   {Sol} H.,  2004, \mn@doi
  [\aap] {10.1051/0004-6361:20040349}, \href
  {https://ui.adsabs.harvard.edu/abs/2004A&A...427..415P} {427, 415}

\bibitem[\protect\citeauthoryear{{Perucho}, {Mart{\'\i}}  \&
  {Quilis}}{{Perucho} et~al.}{2019}]{Perucho2019MNRAS}
{Perucho} M.,  {Mart{\'\i}} J.-M.,   {Quilis} V.,  2019, \mn@doi [\mnras]
  {10.1093/mnras/sty2912}, \href
  {https://ui.adsabs.harvard.edu/abs/2019MNRAS.482.3718P} {482, 3718}

\bibitem[\protect\citeauthoryear{{Rieger}}{{Rieger}}{2019}]{Rieger2019Galax}
{Rieger} F.~M.,  2019, \mn@doi [Galaxies] {10.3390/galaxies7030078}, \href
  {https://ui.adsabs.harvard.edu/abs/2019Galax...7...78R} {7, 78}

\bibitem[\protect\citeauthoryear{{Rieger} \& {Duffy}}{{Rieger} \&
  {Duffy}}{2004}]{Rieger2004}
{Rieger} F.~M.,  {Duffy} P.,  2004, \mn@doi [\apj] {10.1086/425167}, \href
  {https://ui.adsabs.harvard.edu/abs/2004ApJ...617..155R} {617, 155}

\bibitem[\protect\citeauthoryear{{Rieger} \& {Duffy}}{{Rieger} \&
  {Duffy}}{2016}]{Rieger2016}
{Rieger} F.~M.,  {Duffy} P.,  2016, \mn@doi [\apj]
  {10.3847/1538-4357/833/1/34}, \href
  {https://ui.adsabs.harvard.edu/abs/2016ApJ...833...34R} {833, 34}

\bibitem[\protect\citeauthoryear{{Rieger} \& {Duffy}}{{Rieger} \&
  {Duffy}}{2019}]{Rieger2019ApJ}
{Rieger} F.~M.,  {Duffy} P.,  2019, \mn@doi [\apjl] {10.3847/2041-8213/ab563f},
  \href {https://ui.adsabs.harvard.edu/abs/2019ApJ...886L..26R} {886, L26}

\bibitem[\protect\citeauthoryear{{Rieger} \& {Duffy}}{{Rieger} \&
  {Duffy}}{2021}]{Rieger2021}
{Rieger} F.~M.,  {Duffy} P.,  2021, \mn@doi [\apjl] {10.3847/2041-8213/abd567},
  \href {https://ui.adsabs.harvard.edu/abs/2021ApJ...907L...2R} {907, L2}

\bibitem[\protect\citeauthoryear{{Rieger} \& {Duffy}}{{Rieger} \&
  {Duffy}}{2022}]{Rieger2022ApJ}
{Rieger} F.~M.,  {Duffy} P.,  2022, \mn@doi [\apj] {10.3847/1538-4357/ac729c},
  \href {https://ui.adsabs.harvard.edu/abs/2022ApJ...933..149R} {933, 149}

\bibitem[\protect\citeauthoryear{{Rieger}, {Bosch-Ramon}  \& {Duffy}}{{Rieger}
  et~al.}{2007}]{Rieger2007}
{Rieger} F.~M.,  {Bosch-Ramon} V.,   {Duffy} P.,  2007, \mn@doi [\apss]
  {10.1007/s10509-007-9466-z}, \href
  {https://ui.adsabs.harvard.edu/abs/2007Ap&SS.309..119R} {309, 119}

\bibitem[\protect\citeauthoryear{{Rossi}, {Mignone}, {Bodo}, {Massaglia}  \&
  {Ferrari}}{{Rossi} et~al.}{2008}]{Rossi2008A&A}
{Rossi} P.,  {Mignone} A.,  {Bodo} G.,  {Massaglia} S.,   {Ferrari} A.,  2008,
  \mn@doi [\aap] {10.1051/0004-6361:200809687}, \href
  {https://ui.adsabs.harvard.edu/abs/2008A&A...488..795R} {488, 795}

\bibitem[\protect\citeauthoryear{{Rossi}, {Bodo}, {Capetti}  \&
  {Massaglia}}{{Rossi} et~al.}{2017}]{Rossi2017A&A}
{Rossi} P.,  {Bodo} G.,  {Capetti} A.,   {Massaglia} S.,  2017, \mn@doi [\aap]
  {10.1051/0004-6361/201730594}, \href
  {https://ui.adsabs.harvard.edu/abs/2017A&A...606A..57R} {606, A57}

\bibitem[\protect\citeauthoryear{{Scheck}, {Aloy}, {Mart{\'\i}}, {G{\'o}mez}
  \& {M{\"u}ller}}{{Scheck} et~al.}{2002}]{Scheck2002MNRAS}
{Scheck} L.,  {Aloy} M.~A.,  {Mart{\'\i}} J.~M.,  {G{\'o}mez} J.~L.,
  {M{\"u}ller} E.,  2002, \mn@doi [\mnras] {10.1046/j.1365-8711.2002.05210.x},
  \href {https://ui.adsabs.harvard.edu/abs/2002MNRAS.331..615S} {331, 615}

\bibitem[\protect\citeauthoryear{{Sironi}, {Rowan}  \& {Narayan}}{{Sironi}
  et~al.}{2021}]{Sironi2021ApJL}
{Sironi} L.,  {Rowan} M.~E.,   {Narayan} R.,  2021, \mn@doi [\apjl]
  {10.3847/2041-8213/abd9bc}, \href
  {https://ui.adsabs.harvard.edu/abs/2021ApJ...907L..44S} {907, L44}

\bibitem[\protect\citeauthoryear{{Snios} et~al.,}{{Snios}
  et~al.}{2019}]{Snios2019}
{Snios} B.,  et~al., 2019, \mn@doi [\apj] {10.3847/1538-4357/aafaf3}, \href
  {https://ui.adsabs.harvard.edu/abs/2019ApJ...871..248S} {871, 248}

\bibitem[\protect\citeauthoryear{{Sun}, {Yang}, {Rieger}, {Liu}  \&
  {Aharonian}}{{Sun} et~al.}{2018}]{Sun2018}
{Sun} X.-N.,  {Yang} R.-Z.,  {Rieger} F.~M.,  {Liu} R.-Y.,   {Aharonian} F.,
  2018, \mn@doi [\aap] {10.1051/0004-6361/201731716}, \href
  {https://ui.adsabs.harvard.edu/abs/2018A&A...612A.106S} {612, A106}

\bibitem[\protect\citeauthoryear{{Tavecchio}}{{Tavecchio}}{2021}]{Tavecchio2020}
{Tavecchio} F.,  2021, \mn@doi [\mnras] {10.1093/mnras/staa4009}, \href
  {https://ui.adsabs.harvard.edu/abs/2021MNRAS.501.6199T} {501, 6199}

\bibitem[\protect\citeauthoryear{{Tchekhovskoy} \& {Bromberg}}{{Tchekhovskoy}
  \& {Bromberg}}{2016}]{Tchekhovskoy2016MNRAS}
{Tchekhovskoy} A.,  {Bromberg} O.,  2016, \mn@doi [\mnras]
  {10.1093/mnrasl/slw064}, \href
  {https://ui.adsabs.harvard.edu/abs/2016MNRAS.461L..46T} {461, L46}

\bibitem[\protect\citeauthoryear{{Wang}, {Reville}, {Liu}, {Rieger}  \&
  {Aharonian}}{{Wang} et~al.}{2021}]{Wang2021MNRAS}
{Wang} J.-S.,  {Reville} B.,  {Liu} R.-Y.,  {Rieger} F.~M.,   {Aharonian}
  F.~A.,  2021, \mn@doi [\mnras] {10.1093/mnras/stab1458}, \href
  {https://ui.adsabs.harvard.edu/abs/2021MNRAS.505.1334W} {505, 1334}

\bibitem[\protect\citeauthoryear{{Webb}, {Barghouty}, {Hu}  \& {le
  Roux}}{{Webb} et~al.}{2018}]{Webb2018}
{Webb} G.~M.,  {Barghouty} A.~F.,  {Hu} Q.,   {le Roux} J.~A.,  2018, \mn@doi
  [\apj] {10.3847/1538-4357/aaae6c}, \href
  {https://ui.adsabs.harvard.edu/abs/2018ApJ...855...31W} {855, 31}

\bibitem[\protect\citeauthoryear{{Webb}, {Al-Nussirat}, {Mostafavi},
  {Barghouty}, {Li}, {le Roux}  \& {Zank}}{{Webb} et~al.}{2019}]{Webb2019}
{Webb} G.~M.,  {Al-Nussirat} S.,  {Mostafavi} P.,  {Barghouty} A.~F.,  {Li} G.,
   {le Roux} J.~A.,   {Zank} G.~P.,  2019, \mn@doi [\apj]
  {10.3847/1538-4357/ab2fca}, \href
  {https://ui.adsabs.harvard.edu/abs/2019ApJ...881..123W} {881, 123}

\bibitem[\protect\citeauthoryear{{Webb}, {Mostafavi}, {Al-Nussirat},
  {Barghouty}, {Li}, {le Roux}  \& {Zank}}{{Webb} et~al.}{2020}]{Webb2020}
{Webb} G.~M.,  {Mostafavi} P.,  {Al-Nussirat} S.,  {Barghouty} A.~F.,  {Li} G.,
   {le Roux} J.~A.,   {Zank} G.~P.,  2020, \mn@doi [\apj]
  {10.3847/1538-4357/ab8635}, \href
  {https://ui.adsabs.harvard.edu/abs/2020ApJ...894...95W} {894, 95}

\makeatother
\end{thebibliography}


\end{document}